\documentclass[twocolumn,english,aps, prd, amsmath, amssymb, amsfonts, superscriptaddress,eqsecnum, nofootinbib]{revtex4-2}
\usepackage[T1]{fontenc}
\usepackage[utf8]{luainputenc}
\usepackage{amsmath}
\usepackage{amssymb}
\usepackage{graphicx}
\usepackage{esint}

\makeatletter
\usepackage[T1]{fontenc}
\usepackage[utf8]{luainputenc}

\allowdisplaybreaks[4]

\usepackage{float}
\usepackage{wasysym}
\usepackage{braket}

\usepackage{setspace}

\renewcommand{\baselinestretch}{1}
\makeatother

\makeatother

\usepackage{babel}
\begin{document}
\title{Describing the Migdal effect with a bremsstrahlung-like process and
many-body effects}
\author{Zheng-Liang Liang}
\email{liangzl@mail.buct.edu.cn}

\affiliation{College of Mathematics and Physics, Beijing University of Chemical
Technology~\\
Beijing 100029, China}
\author{Chongjie Mo}
\email{cjmo@csrc.ac.cn}

\affiliation{Beijing Computational Science Research Center, Beijing, 100193, China}
\author{Fawei Zheng}
\email{fwzheng@bit.edu.cn}

\affiliation{Centre for Quantum Physics, Key Laboratory of Advanced~\\
 Optoelectronic Quantum Architecture and Measurement(MOE),~\\
School of Physics, Beijing Institute of Technology, Beijing, 100081,
China}
\author{Ping Zhang}
\email{pzhang2012@qq.com}

\affiliation{School of Physics and Physical Engineering, Qufu Normal University~\\
 Qufu, 273165, China}
\affiliation{Institute of Applied Physics and Computational Mathematics~\\
Beijing, 100088, China}
\begin{abstract}
Recent theoretical studies have suggested that the suddenly recoiled
atom struck by dark matter~(DM) particle is much more likely to excite
or lose its electrons than expected. Such Migdal effect opens a new
avenue for exploring the sub-GeV DM particles. There have been various
attempts to describe the Migdal effect in liquid and semiconductor
targets. In this paper we incorporate the treatment of the bremsstrahlung
process and the electronic many-body effects to give a full description
of the Migdal effect in bulk semiconductor targets diamond and silicon.
Compared with the results obtained with the atom-centered localized
Wannier functions~(WFs) under the framework of the tight-binding~(TB)
approximation, the method proposed in this study yields much larger
event rates in the low energy regime, due to a $\omega^{-4}$ scaling.
We also find that the effect of the bremsstrahlung photon mediating
the Coulomb interaction between recoiled ion and the electron-hole
pair is equivalent to that of the exchange of a single phonon. 
\end{abstract}
\maketitle

\section{Introduction}

Although there has been overwhelming evidence for the existence of
the dark matter (DM) from astrophysics and cosmology, its particle
nature still remains mysterious. In the past decades, tremendous efforts
have been invested into the search of the weakly interacting massive
particles (WIMPs), one of the most promising DM candidate, but conclusive
evidence has not yet emerged. Owing to the continuous developments
in detector technologies in recent years, the frontier of the dark
matter direct detection has been pushed to the mass range below the
GeV scale, where the traditional detection methods based on the nuclear
scattering are expected to lose sensitivity. So more and more theorists
and experimentalists have begun to shift to other alternative proposals
based on new detection channels and materials, such as with semiconductors~\citep{Essig:2011nj,Graham:2012su,Essig:2015cda,Hochberg:2016sqx},
Dirac materials~\citep{Hochberg:2017wce,Coskuner:2019odd,Geilhufe:2019ndy},
superconductors~\citep{Hochberg:2015pha,Hochberg:2016ajh}, superfluid
helium~\citep{Knapen2017,Caputo:2019cyg,Caputo:2019xum}, and via
phonon excitations~\citep{Griffin:2018bjn,Knapen2018,Campbell-Deem:2019hdx}
and bremsstrahlung photons~\citep{Kouvaris:2016afs,Bell:2019egg},
as well as other proposals and analyses~\citep{Essig:2012yx,Lee:2015qva,Hochberg:2015fth,Bloch:2016sjj,Derenzo:2016fse,Hochberg:2016ntt,Essig:2016crl,Kadribasic:2017obi,Essig:2017kqs,Arvanitaki:2017nhi,Budnik:2017sbu,SHARMA2017326,Cavoto:2017otc,Liang:2018bdb,Heikinheimo:2019lwg,Trickle:2019nya,Trickle:2019ovy,Catena:2019gfa,Andersson:2020uwc,Trickle:2020oki,Griffin:2020lgd}.

The Migdal effect has aroused wide interest recently because it is
theoretically clarified that the suddenly struck nucleus can produce
ionized electrons more easily than anticipated for an incident sub-GeV
DM particle~\citep{Ibe2018}, so exploring the sub-GeV parameter
space is possible for the present detection technologies. There has
been numerous theoretical proposals~\citep{Dolan:2017xbu,Baxter:2019pnz,PhysRevD.102.043007,GrillidiCortona:2020owp,Liu_2020}
and experimental efforts~\citep{Aprile:2019jmx,Liu:2019kzq,Nakamura:2020kex}
dedicated to detecting sub-GeV DM particles via the Migdal effect.
Ref.~\citep{Essig2019} first investigated the Migdal effect in semiconductor
targets by exploring the connection between the DM-electron scattering
in semiconductors and Migdal processes in isolated atoms. In our previous
study~\citep{PhysRevD.102.043007}, we proposed to describe the Migdal
effect in semiconductors under the framework of the tight-binding
(TB) approximation, where a Galilean boost operator is imposed on
the recoiled ion to account for the highly local impulsive effect
brought by the incident DM particle, while the extensive nature of
the electrons in solids is reflected in the hopping integrals.

Meanwhile, nontrivial collective behavior in condensed matter system
has also attracted attention and has been considered as a possible
origin of observed signal features in various experiments~\citep{Kurinsky_2020}.
Ref.~\citep{Kurinsky_2020} put forward two mechanisms based on the
plasmon production induced by DM particles to explain the existing
signal lineshapes that are at odds with the standard interpretations
of the DM-electron scattering. As a typical many-body phenomenon in
solids, the plasmon excitation cannot be understood in terms of standard
two-body scattering, or non-interacting single-particle states. Therefore,
the many-body physics may shed new light on the interpretation of
the DM-detector interactions.

On the other hand, Ref.~\citep{Kozaczuk:2020uzb} examined the above
postulates by concretely calculating the bremsstrahlung emission of
plasmon induced indirectly by a DM particle. While the many-body effect,
namely, the plasmon resonance can be well described in the same way
as in the electron energy loss spectroscopy (EELS) analysis, the effect
of a fast charged electron transversing the material is replaced with
an abruptly recoiling ion, and this part of physics can be summarized
with the bremsstrahlung-like process, in the context of classical
electrodynamics or quantum mechanics.

In this study we integrate above two aspects, i.e., the electronic
many-body effect and the bremsstrahlung-like description of the recoiled
ion, into a coherent method to describe the Migdal effect in solids.
In this picture the suddenly recoiled ion excites a primary electron-hole
pair via the bremsstrahlung photon, while the many-body physics is
encoded in the dielectric function $\epsilon\left(\mathbf{k},\omega\right)$,
which is responsible for the screening of the pure Coulomb interaction
and the emergence of collective plasma oscillations. It is tempting
to investigate the consequence of the many-body physics on the DM
semiconductor detector. In order to take into account the local field
effect, the calculation of $\epsilon\left(\mathbf{k},\omega\right)$
is extended to the case in crystalline environment using the random
phase approximation (RPA). Based on this method, we concretely estimate
the Migdal excitation event rates in bulk diamond and silicon semiconductors,
respectively. Moreover, considering there have existed other methods
describing the Migdal effect in solids, such as the aforementioned
TB approximation, thus we also make comparisons with the event rates
obtained from the TB approximation, or the atom-centered Wannier functions~(WFs)~\citep{PhysRevD.102.043007}.

This paper is organized as follows. We begin Sec.~\ref{sec:qft}
by giving the quantum mechanical formalism and relevant scheme for
calculating the Migdal excitation event rates induced by DM particles.
Based on this approach, we then specifically calculate the event rates
for diamond and silicon semiconductors, respectively in Sec.~\ref{sec:Computation_Results}.
We conclude and make some comments on the methodology in Sec.~\ref{sec:Conclusions}.
Some details of the derivation in the main text are provided in the
Appendices.

\section{\label{sec:qft}Electronic excitation in the quantum field theory
description}

We begin this section with a short review of the description of the
Migdal effect in crystal structures in the context of the quantum
theory.
\begin{figure*}[t]
\begin{centering}
\includegraphics[scale=0.8]{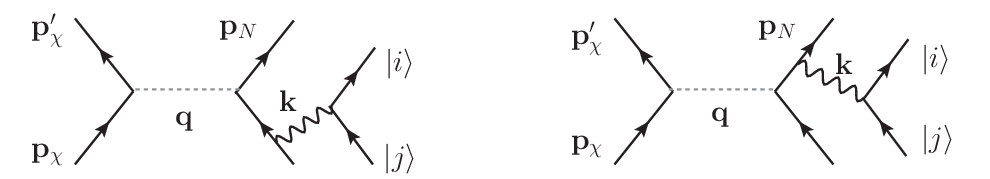}
\par\end{centering}
\caption{\label{fig:Diagrams}The diagrams of the process $\chi\left(p_{\chi}\right)+N\rightarrow\chi\left(p_{\chi}'\right)+N\left(p_{N}\right)+h\left(j\right)+e\left(i\right)$.}
\end{figure*}

\subsection{A general formula for Migdal effect in quantum field theory}

The Migdal effect in solids refers to the process in which the incident
DM particle collides with a nucleus in crystal, and the recoiled nucleus
excites an electron across the band gap from valence state $\ket{j}$
to a conduction state $\ket{i}$. As has been pointed out in our previous
study~\citep{PhysRevD.102.043007} that the boost argument for the
Migdal effect in an isolated atom used in Ref.~\citep{Ibe2018} no
longer applies for the crystalline solids, because one can not regard
the whole crystal target as a big nucleus: it is the struck nucleus
in the crystal that is recoiling, but not the whole crystal. A description
fixed to the crystal is preferred. Inspired by the treatment of the
plasmon production in Ref.~\citep{Kozaczuk:2020uzb}, in this study
we discuss the Migdal effect with similar description in the hope
that sudden acceleration effect of the nucleus will be well captured
with a bremsstrahlung-like process. This process can also be understood
as the two-body scattering between the DM particle and a nucleus,
with the scattered DM particle and the nucleus, as well as the electron-hole
pair as the final states, i.e., the process $\chi\left(p_{\chi}\right)+N\rightarrow\chi\left(p_{\chi}'\right)+N\left(p_{N}\right)+\,$electron-hole
pair, where $\chi$ stands for the DM particle, $N\left(p_{N}\right)$~$\left(N\left(p_{N}'\right)\right)$
is the nucleus before (after) the collision. The relevant Feynman
diagrams are presented in Fig.~\ref{fig:Diagrams}, where $e\left(i\right)$
and $h\left(j\right)$ represent the electronic conduction and valence
states, respectively, from which the amplitude is read as 
\begin{eqnarray}
i\mathcal{M} & = & \sum_{\mathbf{k}}\frac{\left(-i\right)}{V}V_{\chi N}\left(\mathbf{q}\right)\,V_{Ne}\left(\mathbf{k}\right)\:\braket{i|e^{i\mathbf{k}\cdot\hat{\mathbf{x}}}|j}\nonumber \\
 & \times & \left[\frac{1}{\frac{p_{N}^{2}}{2m_{N}}+\varepsilon_{ij}-\frac{\left|\mathbf{p}_{N}+\mathbf{k}\right|^{2}}{2m_{N}}}-\frac{1}{\varepsilon_{ij}+\frac{k^{2}}{2m_{N}}}\right],\label{eq:feynman rules}
\end{eqnarray}
where $\mathbf{q}=-\mathbf{p}_{N}-\mathbf{k}$, $p_{N}=\left|\mathbf{p}_{N}\right|$,
$k=\left|\mathbf{k}\right|$, $V$ is the volume of the crystal, $m_{N}$
is the nucleus mass, $\varepsilon_{ij}=\varepsilon_{i}-\varepsilon_{j}$
is the energy difference between the conduction states $\left\{ \ket{i}\right\} $
and valence states $\left\{ \ket{j}\right\} $, $V_{\chi N}\left(\mathbf{q}\right)$
represents the DM-nucleus contact interaction $V_{\chi N}\left(\mathbf{x}\right)$
in momentum space, which is connected to the DM-nucleus cross section
with the following relation, 
\begin{eqnarray}
\left|V_{\chi N}\left(\mathbf{q}\right)\right|^{2} & = & \left|\int\mathrm{d}^{3}x\,e^{-i\mathrm{\mathbf{q}\cdot\mathbf{x}}}\,V_{\chi N}\left(\mathbf{x}\right)\right|^{2}\nonumber \\
 & = & \frac{A^{2}\pi\sigma_{\chi n}}{\mu_{\chi n}^{2}},
\end{eqnarray}
where $A$ is the atomic number of the target nucleus, $\sigma_{\chi n}$
and $\mu_{\chi n}=m_{n}\,m_{\chi}/\left(m_{n}+m_{\chi}\right)$ represent
the cross section and the reduced mass of the DM-nucleon pair, respectively.
$V_{Ne}\left(\mathbf{k}\right)$ is the ion-electron Coulomb interaction
propagator 
\begin{eqnarray}
V_{Ne}\left(\mathbf{k}\right) & = & \int\mathrm{d}^{3}x\,e^{-i\mathrm{\mathbf{k}\cdot\mathbf{x}}}\,\frac{Z_{\mathrm{ion}}e^{2}}{4\pi\,\left|\mathbf{x}\right|}\nonumber \\
 & = & \frac{Z_{\mathrm{ion}}4\pi\alpha}{k^{2}},
\end{eqnarray}
where $Z_{\mathrm{ion}}$ is the number of valence electrons, and
$\alpha$ is the fine structure constant. In the soft limit where
$\mathbf{p}_{N}\cdot\mathbf{k}/m_{N}\ll\varepsilon_{ij}$ and $\left|\mathbf{k}\right|\ll\left|\mathbf{p}_{N}\right|$,
the nucleus propagator in Eq.~(\ref{eq:feynman rules}) can be simplified
as~\citep{Kozaczuk:2020uzb} 
\begin{eqnarray}
\frac{1}{\frac{p_{N}^{2}}{2m_{N}}+\varepsilon_{ij}-\frac{\left|\mathbf{p}_{N}+\mathbf{k}\right|^{2}}{2m_{N}}}-\frac{1}{\varepsilon_{ij}+\frac{k^{2}}{2m_{N}}} & \simeq & \left(\frac{\mathbf{p}_{N}\cdot\mathbf{k}}{m_{N}}\right)\frac{1}{\varepsilon_{ij}^{2}}.\nonumber \\
\end{eqnarray}
At this stage, one may wonder, whether the bremsstrahlung-like process
(where an electron is excited via exchanging a photon) amounts to
a complete description of the actual physical process, considering
that in discussion of the Migdal effect in an isolated atom, the excitation
process is described with a full evolution of the Hamiltonian. So
if one can show that bremsstrahlung-like description is equivalent
to the boost argument for the Migdal effect in an isolated atom, the
bremsstrahlung-like narrative can be convincingly generalized to the
crystalline solids. To this end, we apply above discussion to an isolated
atom by substituting the valence~($\ket{j}$) and conduction~($\ket{i}$)
states, as well as the valence charge $Z_{\mathrm{ion}}$, with initial~($\ket{\alpha}$),
final ~($\ket{\beta}$) atomic states, and the atomic effective charge
$Z_{\mathrm{eff}}$, respectively, and the amplitude in Fig.~\ref{fig:Diagrams}
can be read off as the following,
\begin{eqnarray}
i\mathcal{M} & = & \left(-i\right)V_{\chi N}\left(\mathbf{q}\right)\,\int\frac{\mathrm{d}^{3}k}{\left(2\pi\right){}^{3}}\frac{Z_{\mathrm{eff}}\,4\pi\alpha}{k^{2}}\:\braket{\beta|e^{i\mathbf{k}\cdot\hat{\mathbf{x}}}|\alpha}\nonumber \\
 &  & \times\frac{\mathbf{v}_{N}\cdot\mathbf{k}}{\left(E_{\beta}-E_{\alpha}\right)^{2}},
\end{eqnarray}
where $\mathbf{v}_{N}=\mathbf{p}_{N}/m_{N}$ represents the velocity
of the recoiled atom, and the correspondence $\frac{1}{V}\sum_{\mathbf{k}}\sim\int\frac{\mathrm{d}^{3}k}{\left(2\pi\right){}^{3}}$
is used. On the other hand, as demonstrated in Ref.~\citep{Knapen:2020aky}
that the excitation rate derived with the boost argument in Ref.~\citep{Ibe2018}
is connected to above bremsstrahlung-like approach through the relation
\begin{eqnarray}
 &  & \braket{\beta|e^{im_{e}\mathbf{v}_{N}\cdot\hat{\mathbf{x}}}|\alpha}\simeq im_{e}\mathbf{v}_{N}\cdot\braket{\beta|\hat{\mathbf{x}}|\alpha}\nonumber \\
 &  & =\frac{Z_{\mathrm{eff}}\,4\pi\alpha}{\left(E_{\beta}-E_{\alpha}\right)^{2}}\int\frac{\mathrm{d}^{3}k}{\left(2\pi\right){}^{3}}\frac{\mathbf{v}_{N}\cdot\mathbf{k}}{k^{2}}\braket{\beta|e^{i\mathbf{k}\cdot\hat{\mathbf{x}}}|\alpha},\label{eq:atom_relation}
\end{eqnarray}
where $m_{e}$ is the electron mass, and $E_{\alpha}$~($E_{\beta}$)
is the energy of the initial~(final) state. Thus the bremsstrahlung-like
framework can also describe the Migdal effect in an isolated atom.
At first sight, one may conclude that this relation also holds in
solids, and hence the atomic formulation can be directly generalized
to the semiconductors~\citep{Essig2019,Liu_2020}. However, this
is not true because the derivation of Eq.~(\ref{eq:atom_relation})
rests on a central Coulomb potential $-Z_{\mathrm{eff}}e^{2}/r$,
while in the crystalline environment the itinerant electrons are also
subjected to the Coulomb interaction from all neighboring ions, and
thus such generalization is invalid. In contrast, our rationale is
that, since the bremsstrahlung-like narrative has been justified in
the case of isolated atoms, we generalize this description to the
case in semiconductors, which leads to Eq.~(\ref{eq:feynman rules}),
rather than $\braket{i|e^{im_{e}\mathbf{v}_{N}\cdot\hat{\mathbf{x}}}|j}$. 

Moreover, if one takes into consideration the screening effect by
introducing the dielectric function $\epsilon^{-1}\left(\mathbf{k},\omega\right)$~\citep{Kurinsky_2020},
the total event rate for the DM flux impinging on a semiconductor
and then exciting a primary electron-hole pair can be written as\footnote{{\selectfont A detailed derivation of this event rate is arranged
in Appendix~\ref{sec:cross_section}}}: 
\begin{eqnarray}
R & = & \frac{\rho_{\chi}}{m_{\chi}}N_{T}\left\langle \sigma v\right\rangle \nonumber \\
 & = & \frac{\rho_{\chi}}{m_{\chi}}\frac{4\pi^{2}A^{2}\sigma_{\chi n}\,Z_{\mathrm{ion}}^{2}\,\alpha\,N_{T}}{3\,\mu_{\chi n}^{2}\,\left(2\pi\right)^{6}}\int\mathrm{d^{3}}v\frac{f_{\chi}\left(\mathbf{v}\right)}{v}\int\mathrm{d}^{3}p_{N}\frac{\left|\mathbf{v}_{\mathrm{ion}}\right|^{2}}{p_{N}}\nonumber \\
 &  & \times\int\frac{\mathrm{d}\omega}{\omega^{4}}\varTheta\left[v-v_{\mathrm{min}}\left(p_{N},\,\omega\right)\right]\int\mathrm{d}^{3}k\frac{2\times4\pi^{2}\alpha}{V\,k^{2}\left|\epsilon\left(\mathbf{k},\omega\right)\right|^{2}}\nonumber \\
 &  & \times\sum_{i,\,j}\left|\braket{i|e^{i\mathbf{k}\cdot\hat{\mathbf{x}}}|j}\right|^{2}\,\delta\left(\varepsilon_{i}-\varepsilon_{j}-\omega\right),\label{eq:eventRate0}
\end{eqnarray}
where the bracket $\left\langle \cdots\right\rangle $ denotes the
average over the DM velocity distribution, $\rho_{\chi}$ represents
the DM local density, $N_{T}$ is number of the nuclei in target,
$f_{\chi}\left(\mathbf{v}\right)$ is the DM velocity distribution,
$\mathbf{v}_{\mathrm{ion}}=\mathbf{p}_{N}/m_{N}$ is the velocity
of the recoiled nucleus, $\omega$ denotes the relevant energy difference,
and $\Theta$ is the Heaviside step function; the factor $1/3$ stems
from the average over all directions of $\mathbf{k}$, given the isotropic
nature of the electron gas; in the last line the factor $2$ counts
the two degenerate spin states. For the given $p_{N}$ and $\omega$,
function $v_{\mathrm{min}}$ determines the minimum kinetically accessible
velocity for the transition:
\begin{eqnarray}
v_{\mathrm{min}}\left(p_{N},\,\omega\right) & = & \frac{p_{N}}{2\,\mu_{\chi N}}+\frac{\omega}{p_{N}},\label{eq:Vmin}
\end{eqnarray}
with $\mu_{\chi N}=m_{N}\,m_{\chi}/\left(m_{N}+m_{\chi}\right)$ being
the reduced mass of the DM-nucleus pair. In practice, we take $\rho_{\chi}=0.3\,\mathrm{GeV/cm^{3}}$,
and the velocity distribution can be approximated as a truncated Maxwellian
form in the Galactic rest frame, i.e., $f_{\chi}\left(\mathbf{v}\right)\propto\exp\left[-\left|\mathbf{v}+\mathbf{v}_{\mathrm{e}}\right|^{2}/v_{0}^{2}\right]\,\Theta\left(v_{\mathrm{esc}}-\left|\mathbf{v}+\mathbf{v}_{\mathrm{e}}\right|\right)$,
with the Earth's velocity $v_{\mathrm{e}}=230\,\mathrm{km/s}$, the
dispersion velocity $v_{0}=220\,\mathrm{km/s}$ and the Galactic escape
velocity $v_{\mathrm{esc}}=544\,\,\mathrm{km/s}$.

\subsection{Migdal effect with random phase approximation}

As illustrated in general expression of the Migdal event rate Eq.~(\ref{eq:eventRate0}),
the dielectric function $\epsilon^{-1}\left(\mathbf{k},\omega\right)$
plays the central role in the estimate, accounting for the many-body
effects. In this paper we calculate the dielectric function with the
random phase approximation (RPA), which corresponds to the Lindhard
formula: 
\begin{widetext}
\begin{eqnarray}
\mathrm{Im}\left[\epsilon\left(\mathbf{k},\omega\right)\right] & \simeq & 2\times\frac{4\pi^{2}\alpha}{V\,k^{2}}\,\sum_{i,\,j}\left|\braket{i|e^{i\mathbf{k}\cdot\hat{\mathbf{x}}}|j}\right|^{2}\,\delta\left(\varepsilon_{i}-\varepsilon_{j}-\omega\right),
\end{eqnarray}
with which the total event rate including the electronic many-body
effect is recast as follows

\begin{eqnarray}
R & = & \frac{\rho_{\chi}}{m_{\chi}}\frac{4\pi^{2}A^{2}\sigma_{\chi n}\,Z_{\mathrm{ion}}^{2}\,\alpha\,N_{T}}{3\,\mu_{\chi n}^{2}\,\left(2\pi\right)^{6}}\int\mathrm{d^{3}}v\frac{f_{\chi}\left(\mathbf{v}\right)}{v}\int\mathrm{d}^{3}p_{N}\frac{\left|\mathbf{v}_{\mathrm{ion}}\right|^{2}}{p_{N}}\int\frac{\mathrm{d}\omega}{\omega^{4}}\varTheta\left[v-v_{\mathrm{min}}\left(p_{N},\,\omega\right)\right]\int\mathrm{d}^{3}k\,\mathrm{Im}\left[\frac{-1}{\epsilon\left(\mathbf{k},\omega\right)}\right].\label{eq:eventRate1}
\end{eqnarray}
This formula applies for the textbook model for the homogeneous electron
gas. In crystalline structure the translational symmetry for spacetime
reduces to that for the periodic crystal lattice. In this case, above
expression is rewritten as the following\footnote{{\selectfont For more details of the calculation, see Appendix~\ref{sec:cross_section}}}:
\begin{eqnarray}
R & = & \frac{\rho_{\chi}}{m_{\chi}}\frac{4\pi^{2}A^{2}\sigma_{\chi n}\,Z_{\mathrm{ion}}^{2}\,\alpha\,N_{T}}{3\,\mu_{\chi n}^{2}\,\left(2\pi\right)^{3}\Omega}\int\mathrm{d^{3}}v\,\frac{f_{\chi}\left(\mathbf{v}\right)}{v}\int\mathrm{d}^{3}p_{N}\frac{\left|\mathbf{v}_{\mathrm{ion}}\right|^{2}}{p_{N}}\int\frac{\mathrm{d}\omega}{\omega^{4}}\,\varTheta\left[v-v_{\mathrm{min}}\left(p_{N},\,\omega\right)\right]\mathcal{F}\left(\omega\right)\nonumber \\
 & = & \frac{\rho_{\chi}}{m_{\chi}}\frac{2\,A^{2}\sigma_{\chi n}\,Z_{\mathrm{ion}}^{2}\,\alpha\,N_{T}}{3\,\mu_{\chi n}^{2}\,\Omega\,m_{N}^{2}}\int\mathrm{d^{3}}v\,\frac{f_{\chi}\left(\mathbf{v}\right)}{v}\int p{}_{N}^{3}\,\mathrm{d}p_{N}\int\frac{\mathrm{d}\omega}{\omega^{4}}\,\varTheta\left[v-v_{\mathrm{min}}\left(p_{N},\,\omega\right)\right]\mathcal{F}\left(\omega\right),\label{eq:eventRate}
\end{eqnarray}
where the nondimensional factor 

\begin{eqnarray}
\mathcal{F}\left(\omega\right) & = & \sum_{\mathbf{G},\mathbf{G}'}\int_{1\mathrm{BZ}}\frac{\varOmega\,\mathrm{d}^{3}\left[k\right]}{\left(2\pi\right)^{3}}\,\frac{\left(\left[\mathbf{k}\right]+\mathbf{G}\right)\cdot\left(\left[\mathbf{k}\right]+\mathbf{G}'\right)}{\left|\left[\mathbf{k}\right]+\mathbf{G}\right|\left|\left[\mathbf{k}\right]+\mathbf{G}'\right|}\,\mathrm{Im}\left[-\widetilde{\epsilon}_{\mathbf{G},\mathbf{G}'}^{-1}\left(\mathbf{\left[k\right]},\omega\right)\right]\label{eq:crystal form factor}
\end{eqnarray}
represents the averaged energy loss function, with $\Omega$ being
the volume of the unit cell. The transferred momentum $\mathbf{k}$
in Eq.~(\ref{eq:eventRate1}) is expressed uniquely as the sum of
a reciprocal lattice vector $\mathbf{G}$, and corresponding reduced
momentum $\left[\mathbf{k}\right]$ confined in the first BZ, i.e.,
$\mathbf{k}=\left[\mathbf{k}\right]+\mathbf{G}$. $\mathcal{F}\left(\omega\right)$
can be obtained from inverting the microscopic dielectric matrix 

\begin{eqnarray}
\widetilde{\epsilon}_{\mathbf{G},\mathbf{G}'}\left(\mathbf{\left[\mathbf{k}\right]},\omega\right) & = & \delta_{\mathbf{G},\mathbf{G}'}-\frac{1}{V}\frac{4\pi\alpha}{\left|\mathbf{\left[\mathbf{k}\right]}+\mathbf{G}\right|\left|\mathbf{\left[\mathbf{k}\right]}+\mathbf{G}'\right|}\sum_{i,j}\frac{\braket{i|e^{i\left(\mathbf{\mathbf{\mathbf{\left[\mathbf{k}\right]}+\mathbf{G}}}'\right)\cdot\hat{\mathbf{x}}}|j}\braket{j|e^{-i\left(\mathbf{\mathbf{\left[\mathbf{k}\right]+\mathbf{G}}}\right)\cdot\hat{\mathbf{x}}}|i}}{\varepsilon_{ij}-\omega-i0^{+}}\left(n_{i}-n_{j}\right),\label{eq:formFactor}
\end{eqnarray}
where $n_{i}$~($n_{j}$) denotes the occupation number of the state
$\ket{i}$~($\ket{j}$). The non-vanishing $\mathbf{G}$-vectors
in the microscopic dielectric matrix reflect the variation of the
microscopic field over a unit cell. For simplicity, the crystal structure
is still approximated as isotropic in discussion. If one presumes
that the fine structure constant $\alpha$ well suppresses the matrix
elements, and then takes only the terms up to the first order in the
resolvent $\left(I-M\right)^{-1}=I+M+M^{2}+\cdots$, where the identity
matrix $I$ and $M$ represent the first and the the second terms
in Eq.~(\ref{eq:formFactor}), respectively, the inverse matrix can
be approximated as 
\begin{eqnarray}
\tilde{\epsilon}_{\mathbf{G},\mathbf{G}'}^{-1}\left(\mathbf{\mathbf{\left[\mathbf{k}\right]}},\omega\right) & \simeq & \delta_{\mathbf{G},\mathbf{G}'}+\frac{1}{V}\frac{4\pi\alpha}{\left|\mathbf{\left[\mathbf{k}\right]}+\mathbf{G}\right|\left|\mathbf{\left[\mathbf{k}\right]}+\mathbf{G}'\right|}\sum_{i,j}\frac{\braket{i|e^{i\left(\mathbf{\mathbf{\mathbf{\left[\mathbf{k}\right]}+\mathbf{G}}}'\right)\cdot\hat{\mathbf{x}}}|j}\braket{j|e^{-i\left(\mathbf{\mathbf{\mathbf{\left[\mathbf{k}\right]}+\mathbf{G}}}\right)\cdot\hat{\mathbf{x}}}|i}}{\varepsilon_{ij}-\omega-i0^{+}}\left(n_{i}-n_{j}\right),
\end{eqnarray}
and hence 
\begin{eqnarray}
\mathrm{Im}\left[-\tilde{\epsilon}_{\mathbf{G},\mathbf{G}'}^{-1}\left(\mathbf{\mathbf{\left[\mathbf{k}\right]}},\omega\right)\right] & \simeq & \frac{2}{V}\frac{4\pi^{2}\alpha}{\left|\mathbf{\left[\mathbf{k}\right]}+\mathbf{G}\right|\left|\mathbf{\left[\mathbf{k}\right]}+\mathbf{G}'\right|}\,\sum_{i,\,j}\braket{i|e^{i\left(\mathbf{\mathbf{\mathbf{\left[\mathbf{k}\right]}+\mathbf{G}}}'\right)\cdot\hat{\mathbf{x}}}|j}\braket{j|e^{-i\left(\mathbf{\mathbf{\mathbf{\left[\mathbf{k}\right]}+\mathbf{G}}}\right)\cdot\hat{\mathbf{x}}}|i}\,\delta\left(\varepsilon_{i}-\varepsilon_{j}-\omega\right),
\end{eqnarray}
which exactly corresponds to the case without the screening effect.
In this study the inverse matrix is calculated in a straightforward
way.
\begin{figure*}[t]
\begin{centering}
\includegraphics[scale=0.6]{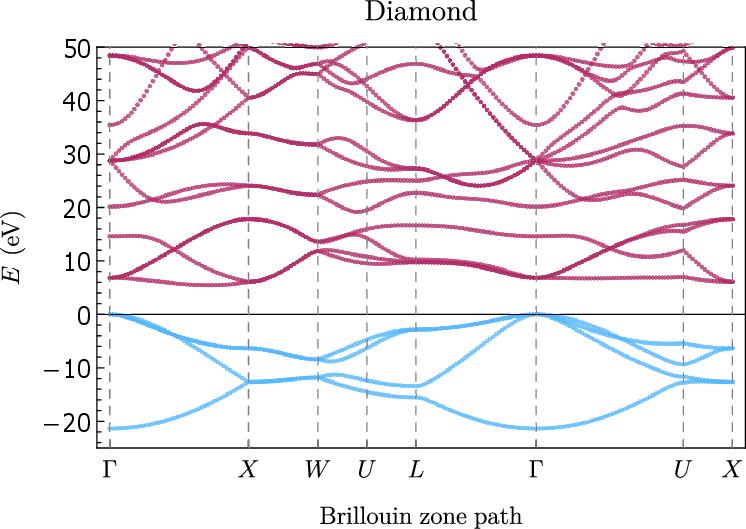}\hspace{0.5cm}\includegraphics[scale=0.6]{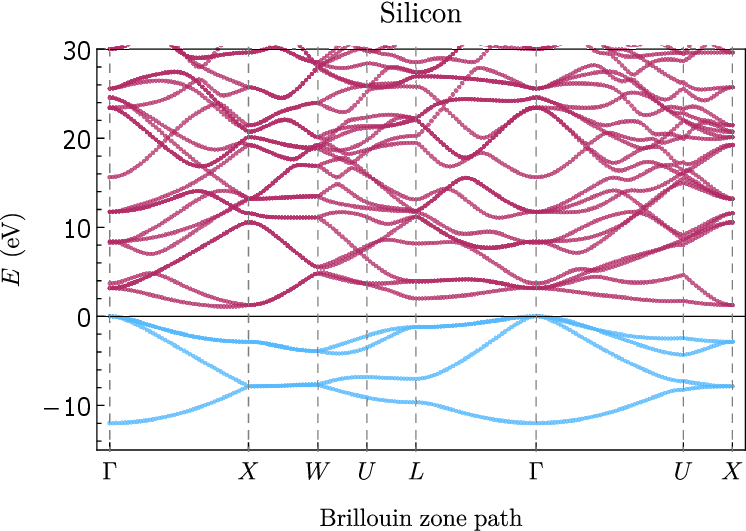}\vspace{0.2cm}
\par\end{centering}
\begin{centering}
\includegraphics[scale=0.85]{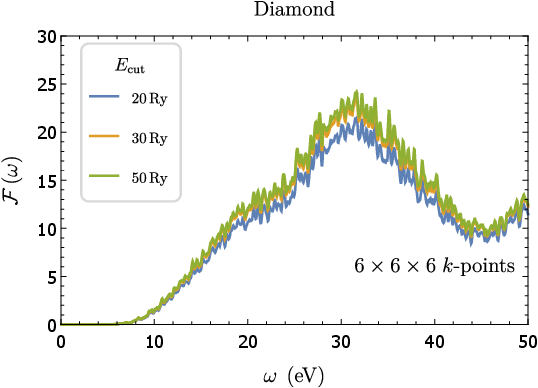}\hspace{0.5cm}\includegraphics[scale=0.85]{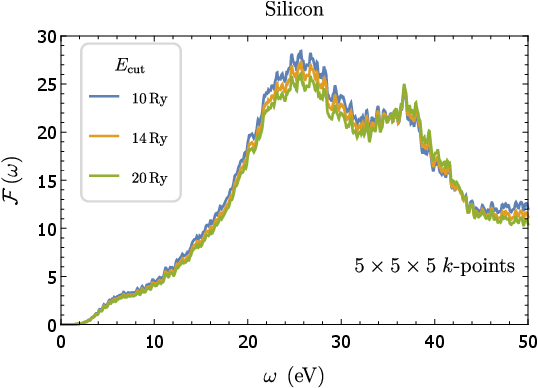}\vspace{0.2cm}
\par\end{centering}
\begin{centering}
\includegraphics[scale=0.85]{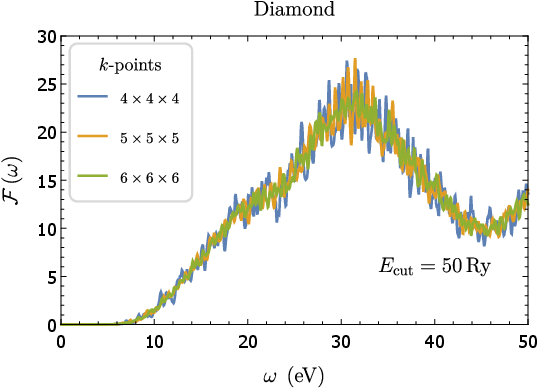}\hspace{0.5cm}\includegraphics[scale=0.85]{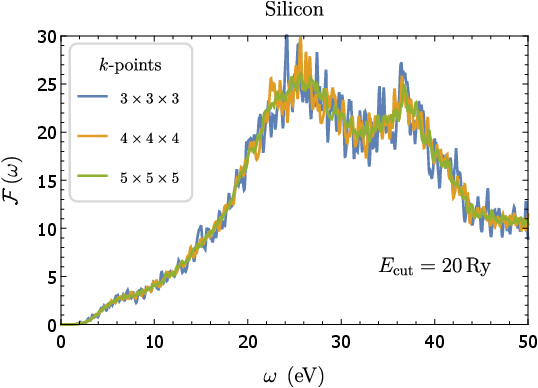}
\par\end{centering}
\caption{\textit{\label{fig:Band}Top: }The scissor-corrected band structure
of bulk diamond~(\textit{left}) and silicon~(\textit{right}) obtained
from DFT calculation, with the band gap $E_{g}=5.47\,\mathrm{eV}$~($1.12\,\mathrm{eV}$).
\textit{Middle}: Given the 6$\times$6$\times$6~(5$\times$5$\times$5)
$k$-point mesh for diamond~(\textit{left}) and silicon~(\textit{right}),
the energy loss spectra $\mathcal{F}\left(\omega\right)$ introduced
in Eq.~(\ref{eq:formFactor}), for $E_{\mathrm{cut}}=10\,\mathrm{Ry}$~($6\,\mathrm{Ry}$),
$20\,\mathrm{Ry}$~($10\,\mathrm{Ry}$), and $50\,\mathrm{Ry}$~($20\,\mathrm{Ry}$),
respectively. \textit{Bottom}: Given $E_{\mathrm{cut}}=50\,\mathrm{Ry}$~($20\,\mathrm{Ry}$)
for diamond~(\textit{left}) and silicon~(\textit{right}), the spectrum
$\mathcal{F}\left(\omega\right)$ for the 6$\times$6$\times$6~(5$\times$5$\times$5),
5$\times$5$\times$5~(4$\times$4$\times$4), and 4$\times$4$\times$4~(3$\times$3$\times$3)
$k$-point mesh, respectively. See text for details.}
\end{figure*}
\end{widetext}

\section{\label{sec:Computation_Results}Computational details and results}

Now we are in a position to put into practice the estimate of the
Migdal excitation event rate. With $\mathtt{Quantum\:Espresso}$ package~\citep{Giannozzi_2009}
plus a norm-conserving pseudopotential~\citep{PhysRevLett.43.1494},
we perform the density functional theory~(DFT) calculation to obtain
the Bloch eigenfunctions and eigenvalues using the local-density approximation~\citep{PhysRevB.23.5048}
for the exchange-correlation functional, on a uniform $6\times6\times6$~($5\times5\times5$)
$k$-point mesh for diamond (silicon) via the Monkhorst-Pack~\citep{PhysRevB.13.5188}
scheme. A core cutoff radius of $1.3\,\mathrm{Bohr}$~($1.8\,\mathrm{Bohr}$)
is adopted and the outermost four electrons are treated as valence
for diamond~(silicon). The energy cut $\varepsilon_{\mathrm{cut}}$
is set to $200\,\mathrm{Ry}$~($70\,\mathrm{Ry}$) and lattice constant
3.577~Å~~(5.429~Å) for diamond (silicon) obtained from experimental
data is adopted. The band structure of diamond (silicon) crystal is
presented in the upper left (right) panel of Fig.~\ref{fig:Band},
where the scissor correction is conducted to match the experimental
value of the band gap $E_{g}=5.47\,\mathrm{eV}$ ($1.12\,\mathrm{eV}$).
\begin{figure*}
\begin{centering}
\includegraphics[scale=0.65]{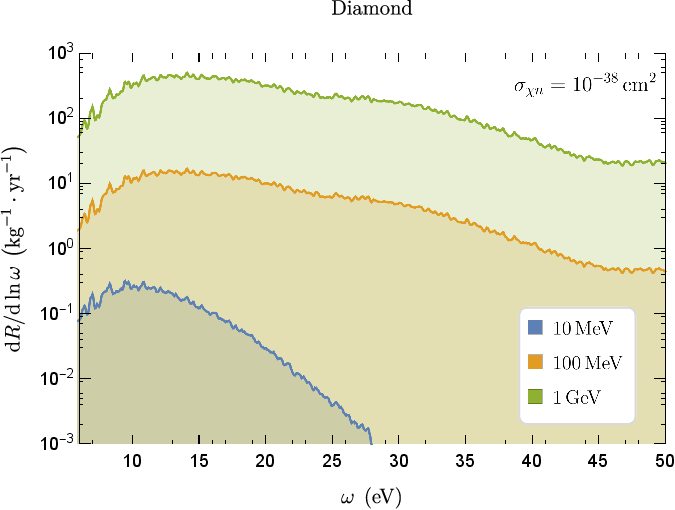}\hspace{0.5cm}\includegraphics[scale=0.65]{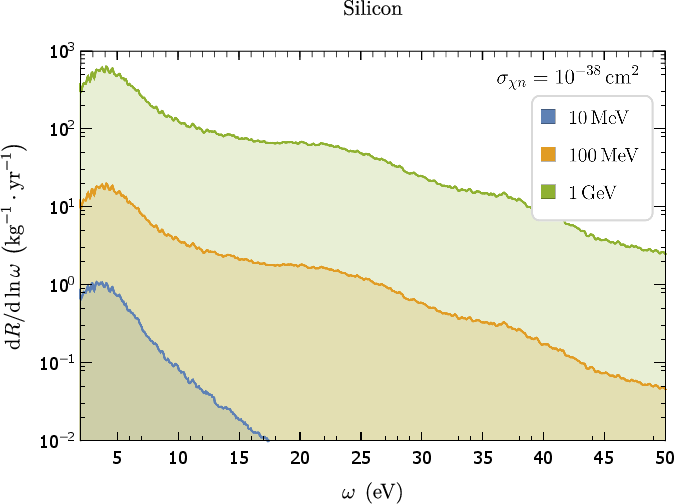}\vspace{0.2cm}
\par\end{centering}
\begin{centering}
\includegraphics[scale=0.68]{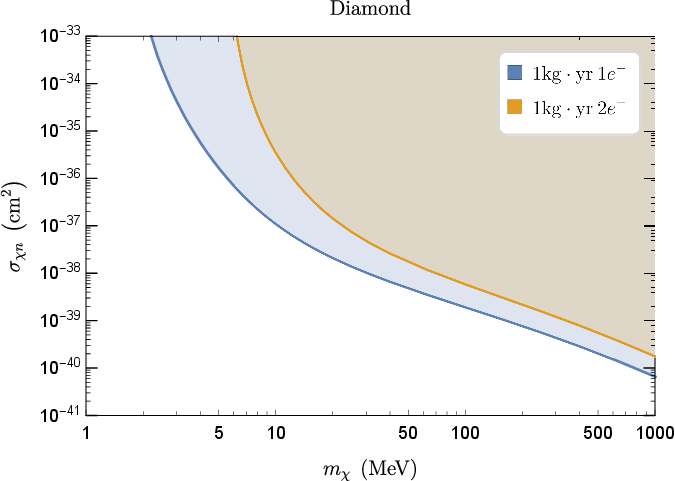}\hspace{0.5cm}\includegraphics[scale=0.68]{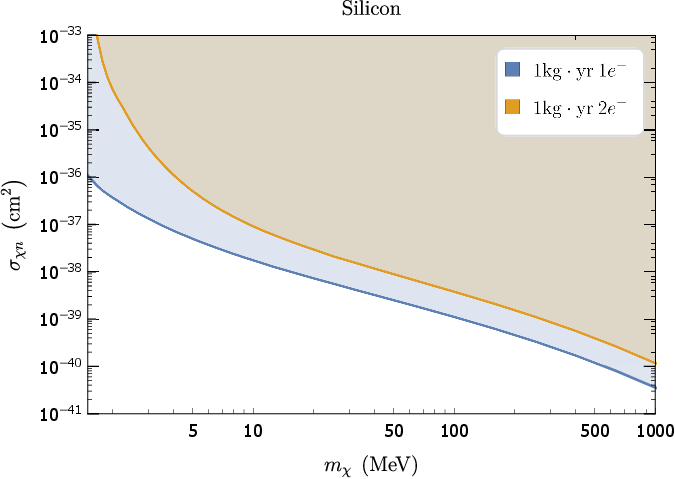}
\par\end{centering}
\caption{\label{fig:Energy Spectrum}\textit{Top:} The differential Migdal
electronic excitation event rate in crystalline diamond~(\textit{left})
and silicon~(\textit{right}), calculated at the RPA level for the
reference values $\sigma_{\chi n}=10^{-38}\,\mathrm{cm}^{2}$ and
$m_{\chi}=10\,\mathrm{MeV}$ (\textit{blue}), $100\,\mathrm{MeV}$(\textit{orange})
and $1\,\mathrm{GeV}$ (\textit{green}), respectively. \textit{Bottom}:
Relevant cross-section sensitivities for the Migdal effect at 90\%
C.L. for a 1 kg-yr diamond~(\textit{left}) and silicon~(\textit{right})
detector, based on the single-electron (\textit{blue}) and the two-electron
(\textit{orange}) bin data, respectively. See text for details.}
\end{figure*}

The matrix $\epsilon_{\mathbf{G},\mathbf{G}'}^{-1}$ is calculated
via directly inverting the matrix Eq.~(\ref{eq:formFactor}) with
the $\mathtt{YAMBO}$ code~\citep{2009CoPhC.180.1392M,Sangalli_2019},
with a smaller matrix cutoff $E_{\mathrm{cut}}$ of $50\,\mathrm{Ry}$~($20\,\mathrm{Ry}$)
for diamond (silicon). An energy bin width $\Delta\omega=0.05\,\mathrm{eV}$
is adopted within the range from $0$ to $50\,\mathrm{eV}$. In the
$\mathtt{YAMBO}$ implementation, the output of the Bloch wavefunctions
are formatted in the form of periodic wavefunctions $\left\{ u_{i\mathbf{k}}\left(\mathbf{x}\right)\right\} $,
normalized within a unit cell, with which the matrix element in Eq.~(\ref{eq:formFactor})
is explicitly written as
\begin{eqnarray}
\braket{i'\,\mathbf{k}'|e^{i\left(\mathbf{k'-\mathbf{k}+\mathbf{G}}\right)\cdot\hat{\mathbf{x}}}|i\,\mathbf{k}} & = & \int_{\Omega}\mathrm{d}^{3}x\,u_{i'\mathbf{k}'}^{*}\left(\mathbf{x}\right)\,e^{i\mathbf{\mathbf{G}}\cdot\mathbf{x}}\,u_{i\mathbf{k}}\left(\mathbf{x}\right),\nonumber \\
\end{eqnarray}
where the integral is performed over the unit cell. This term can
be understood as the Fourier transformation of the squared term $u_{i'\mathbf{k}'}^{*}\left(\mathbf{x}\right)\,u_{i\mathbf{k}}\left(\mathbf{x}\right)$
within the unit cell, and hence the calculation is practically performed
using the discrete fast Fourier transformation~(FFT) technique. So
the resolution in the position space is associated with the truncation
radius of the reciprocal $\mathbf{G}$-vectors in Eq.~(\ref{eq:crystal form factor}),\textit{
$G$,} which is determined from the energy cut $\varepsilon_{\mathrm{cut}}$
through the relation $G=\sqrt{2m_{e}\varepsilon_{\mathrm{cut}}}$.

In practical evaluation of dielectric matrix Eq.~(\ref{eq:formFactor}),
a small broadening parameter $\eta=0.1\,\mathrm{eV}$ is adopted for
both diamond and silicon, instead of an infinitesimal energy width
$0^{+}$. Theoretically, the smaller the parameter $\eta$, the more
accurate the computation is, but on the other hand, a smaller $\eta$
in turn requires a finer energy width $\Delta\omega$ and a denser
$k$-point mesh to smear the spectra. Thus such choice of parameter
$\eta$ is the result of a balance between accuracy, efficiency, and
smoothness. The integrals of the continuous $k$-points in the first
BZ are replaced by the summations over a uniform discrete mesh of
representative $k$-points as follows: 
\begin{eqnarray}
\int_{\mathrm{1BZ}}\frac{\varOmega\,\mathrm{d}^{3}k}{\left(2\pi\right)^{3}}\left(\cdots\right) & \rightarrow & \frac{1}{N_{k}}\sum_{\mathbf{k}}^{N_{k}}\left(\cdots\right),
\end{eqnarray}
with $N_{k}$ being the number of $k$-points sampled in the first
BZ. As mentioned above, a homogeneous set of $6\times6\times6$~($5\times5\times5$)
$k$-points for diamond (silicon) is used in this study. Given these
$k$-point meshes we compute the averaged energy loss functions $\mathcal{F}\left(\omega\right)$
defined in Eq.~(\ref{eq:crystal form factor}) for various choices
of energy cuts for diamond~(left) and silicon~(right) in the middle
row of Fig.~\ref{fig:Band}, while in the third row of Fig.~\ref{fig:Band}
we show the calculated spectra $\mathcal{F}\left(\omega\right)$ for
various $k$-point meshes for the given energy cuts. Following the
convention in computational condensed matter physics, we quantify
the uncertainties in our computation in terms of the tendency of convergence
depending on these $k$-points and cut-energies $E_{\mathrm{cut}}$
adopted in calculation in Fig.~\ref{fig:Band}. In our computation,
the differences between the event rates (after integrating over 0~eV
to 50~eV) calculated from two sets of parameters are found to be
convergent within 5\% for both diamond and silicon. Therefore the
computational parameters for the $k$-point meshes~($6\times6\times6$
for diamond and $5\times5\times5$ for silicon) and the energy cuts~($50\,\mathrm{Ry}$
for diamond and $20\,\mathrm{Ry}$ for silicon) are sufficient to
give a quantitative description of the Migdal effect at the present
stage. 

In the upper panel of Fig.~\ref{fig:Energy Spectrum} shown are the
velocity-averaged energy spectra of the Migdal excitation for DM mass
$m_{\chi}=10\,\mathrm{MeV}$ (blue), $100\,\mathrm{MeV}$(orange)
and $1\,\mathrm{GeV}$ (green), respectively, for a benchmark cross
section $\sigma_{\chi n}=10^{-38}\,\mathrm{cm}^{2}$. Due to a long
tail brought by the non-vanishing $\eta$, which is further amplified
by the factor $\propto\omega^{-4}$, the spectra do not exactly terminate
at the band gaps, so we truncate the spectra at values slightly higher
than the lower edges of their conduction bands. Also due to the dependence
of the $\propto\omega^{-4}$ factor, the spectrum features a peak
roughly at $13\,\mathrm{eV}$ ($5\,\mathrm{eV}$) for diamond (silicon),
and suffers a suppression at the high energy end, exhibiting a drastic
variation in the whole energy range. The energy spectra are calculated
up to $\omega=50\,\mathrm{eV}$, a value falls short to span all the
energy range relevant for ionization signal for DM masses larger than
$100\,\mathrm{MeV}$, considering that the relevant spectra extend
out beyond the energy window. However, in this energy window, one
can still determine or constrain the DM particle parameters from the
single- and two-electrons bins via the Migdal scattering. To fulfill
this purpose, we adopt the model~\citep{Essig:2015cda} where the
extra electron-hole pairs triggered by the primary pair are described
with the mean energy per electron-hole pair $\varepsilon$ in high
energy recoils. In this picture, the ionization charge $Q$ is then
given by 
\begin{eqnarray}
Q\left(\omega\right) & = & 1+\left\lfloor \left(\omega-E_{g}\right)/\varepsilon\right\rfloor ,
\end{eqnarray}
where $\left\lfloor x\right\rfloor $ rounds $x$ down to the nearest
integer. Thus, from the energy spectra we estimate the sensitivity
of a 1 kg-yr detector in the bottom panel of Fig.~\ref{fig:Energy Spectrum},
assuming an average energy $\varepsilon=13\,\mathrm{eV}$~($3.6\,\mathrm{eV}$)
for producing one electron-hole pair for diamond~\citep{Kurinsky:2019pgb}~(silicon~\citep{Essig:2015cda}).
The 90\% C.L. upper limits on DM-nucleon cross section for both a
single-electron (blue) and a two-electron (orange) bins are presented
with no background event assumed.

It should also be noted that our calculation of electron-hole production
rate also includes the contributions from the bremsstrahlung plasmon
production, and we assume that the plasmons decay dominantly to the
electron-hole pairs~\citep{Kozaczuk:2020uzb}.
\begin{figure}
\centering{}\includegraphics[scale=0.65]{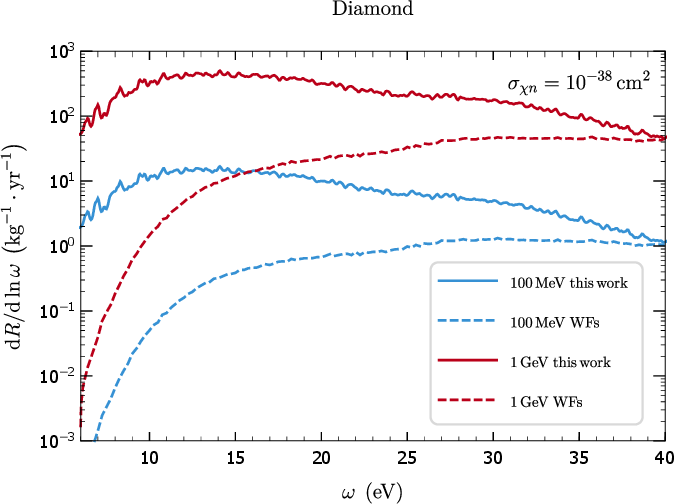}\caption{\label{fig:comparison}A comparison between the Migdal excitation
event rates calculated in this work~(in \textit{solid}) and obtained
in Ref.~\citep{PhysRevD.102.043007} with the localized WFs~(in
\textit{dashed}) in crystal diamond, where the red lines represent
the event rates of a $1\,\mathrm{GeV}$ DM, and the blues represent
the rates of a $100\,\mathrm{MeV}$ DM particle, for the benchmark
cross section $\sigma_{\chi n}=10^{-38}\,\mathrm{cm}^{2}$. See text
for details.}
\end{figure}

\section{\label{sec:Conclusions}Summary and conclusions}

In this paper we make an alternative attempt to describe the Migdal
effect in semiconductor targets, in which the description of the Migdal
effect separates into two parts. Firstly, we incorporate the bremsstrahlung-like
process to account for the drag force exerted on the electrons by
the suddenly struck ion, where the Coulomb interaction is mediated
by the bremsstrahlung photons. Secondly, many-body effects such as
screening, collective behavior are also taken into consideration in
the calculation of excitation rate of the electron-hole pairs. Based
on this method, the Migdal excitation event rates for diamond and
silicon semiconductors are calculated.

As mentioned in previous section, the spectra are modulated by the
factor $\omega^{-4}$, which results in a significant boost to the
excitation event rate in the low energy regime and a suppression towards
the high energy end. Such phenomenon is especially remarkable for
a target possessing a small band gap, such as silicon presented in
right panel of Fig.~\ref{fig:Energy Spectrum}. It is interesting
to make a comparison between the event rates calculated in this work
and the ones obtained with the localized WFs in Ref.~\citep{PhysRevD.102.043007}.
For this purpose we present the two spectra for diamond crystal in
Fig.~\ref{fig:comparison}. Compared to the spectra calculated with
the localized WFs in Ref.~\citep{PhysRevD.102.043007}~(in dashed),
the event rates calculated in this study~(in solid) are found to
be significantly larger in the low energy range near the band gap,
and turn moderately higher towards the high energy end. Apparently,
the $\omega^{-4}$ scaling behavior has not been reflected in the
crystal form factor $\mathcal{F}\left(q,\,E_{e}\right)$ calculated
in Ref.~\citep{PhysRevD.102.043007}.
\begin{figure}[t]
\begin{centering}
\includegraphics[scale=0.65]{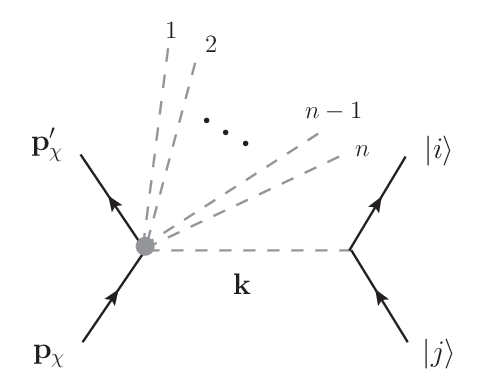}
\par\end{centering}
\caption{\label{fig:phonon-bundle}The diagram for the multiphonon process
where the DM particle excites an electron-hole pair via the exchange
of a single phonon, while most of the transferred momentum of the
DM particle is taken by a bunch of $n$ phonons. See text for details.}
\end{figure}

Finally, we try to interpret this method in the context of the quantization
of the crystal vibration. We begin with the diagram in Fig.~\ref{fig:phonon-bundle}
that represents a process where a pair of electron and hole is excited
via the exchange of single phonon, along with a bunch of phonons created
by the scattering of the DM particle and nucleus. It is noted that
the phonon propagator can be approximated as $iD_{\mathbf{k},\alpha}=i/\left(\varepsilon_{ij}^{2}-\omega_{\mathbf{k},\alpha}^{2}\right)\simeq i/\varepsilon_{ij}^{2}$,
because the phonon eigenfrequncy of branch $\alpha$ at transferred
momentum $\mathbf{k}$, $\omega_{\mathbf{k},\alpha}\sim\mathcal{O}\left(10^{-1}\mathrm{eV}\right)$,
is much smaller than the band gap. As a result of this approximation,
the sum over the products of the two vertices relevant for the excitation
can be contracted as the following, 
\begin{eqnarray}
-\sum_{\alpha}\left(\frac{\boldsymbol{\epsilon}_{\mathbf{k},\alpha}\cdot\mathbf{q}}{\sqrt{m_{N}}}\right)\left(\frac{\boldsymbol{\epsilon}_{\mathbf{k},\alpha}\cdot\mathbf{k}}{\sqrt{m_{N}}}\right) & \simeq & \frac{\mathbf{p}_{N}\cdot\mathbf{k}}{m_{N}},\label{eq:contraction}
\end{eqnarray}
where $\boldsymbol{\epsilon}_{\mathbf{k},\alpha}$ is the phonon eigenvector
of momentum $\mathbf{k}$. Taking into consideration also the interactions
$V_{\chi N}$ and $V_{Ne}$ from the two vertices, respectively, as
well as other necessary factors, one observes that the processes in
Fig.~\ref{fig:Diagrams} and Fig.~\ref{fig:phonon-bundle} are equivalent,
except for a difference in interpreting the recoil effects in the
crystal: in Fig.~\ref{fig:Diagrams} the final state is the recoiled
nucleus, whereas in Fig.~\ref{fig:phonon-bundle} the final states
are a large number of $n$ phonons. In fact, it is implied through
an isotropic harmonic oscillator toy model that in the limit $q\rightarrow\infty$,
the effects of the multiphonon final states can be well summarized
with a recoiled nucleus, which is namely the impulse approximation~\citep{Schober2014}.
In this regard, the two descriptions are essentially identical: the
bremsstrahlung-like process in the soft limit is equivalent to a multiphonon
process in the impulse approximation, where a valence electron is
excited across the band gap via the exchange of a single phonon.

However, in the low momentum transfer regime~(or equivalently, in
the low DM mass regime), the impulse approximation may no longer be
valid, and the recoiling effect should be accounted for with a multiphonon
process. A detailed analysis in Ref.~\citep{Knapen:2020aky} indicates
that for $m_{\chi}\leq50\,\mathrm{MeV}$ the impulse approximation
ceases to be reliable for silicon and germanium targets, where phonons,
rather than a free nucleus, are more appropriate for the description
of the recoiling effect. \vspace{0.4cc}

\textbf{\textit{Note added}} In the final phase of preparing this
paper, which was enlightened by Ref.~\citep{Kozaczuk:2020uzb}, Ref.~\citep{Knapen:2020aky}
appeared, which calculated the Migdal effect in semiconductor targets
silicon and germanium with the similar techniques, and provided heuristic
explanation for the treatment of the bremsstrahlung-like process,
while in this work we calculate the event rates for diamond and silicon.
Ref.~\citep{Knapen:2020aky} also pointed out the importance of the
contribution of the off-diagonal terms of the dielectric matrix to
the total Migdal event rate, which motivates us to further take into
account the off-diagonal terms for a complete calculation of the Migdal
event rates in diamond and silicon targets. 
\begin{acknowledgments}
This work was partly supported by Science Challenge Project under
No.~TZ2016001, by the National Key R\&D Program of China under Grant
under No.~2017YFB0701502, and by National Natural Science Foundation
of China under No.~11625415. C.M. was supported by the NSFC under
Grants No.~12005012, No.~11947202, and No.~U1930402, and by the
China Postdoctoral Science Foundation under Grants No.~2020T130047
and No.~2019M660016. \onecolumngrid 
%\numberwithin{equation}{section} 
%\renewcommand{\theequation}{\thesection\arabic{equation}}
\end{acknowledgments}

\appendix

\section{Excitation event rate in the quantum field theory}

In this Appendix, we provide some detailed derivation of the formulas
in the main text.

\subsection{Feynman rules}

In order to describe the scattering processes in solids it is necessary
to derive the Feynman rules at zero and finite temperatures. Here
a brief review is arranged. We begin with the propagators in the position
space. 
\begin{itemize}
\item The DM-nucleus propagator between space-time coordinates $x$ and
$x'$ is
\begin{eqnarray}
iS_{\chi N}\left(x,\,x'\right) & = & \sum_{\mathbf{q}}\int iV_{\chi N}\left(\mathbf{q}\right)\frac{e^{i\mathbf{q}\cdot\left(\mathbf{x}-\mathbf{x}'\right)}}{V}\frac{e^{-i\omega\left(t_{x}-t_{x'}\right)}\mathrm{d}\omega}{2\pi}\nonumber \\
 & = & \sum_{\mathbf{q}}\int iV_{\chi N}\left(\mathbf{q}\right)\frac{e^{i\mathbf{q}\cdot\left(\mathbf{x}-\mathbf{x}'\right)}}{V}\times\delta\left(t_{x}-t_{x'}\right),
\end{eqnarray}
in which one neglects the retardation of the interaction in the non-relativistic
limit. 
\item The nucleus propagator between $x$ and $x'$ is
\begin{eqnarray}
iS_{NN}\left(x,\,x'\right) & = & \sum_{\mathbf{\mathbf{p}}}\int\left(\frac{i}{\omega-\varepsilon_{\mathbf{\mathbf{p}}}+i0^{+}}\right)\frac{e^{i\mathbf{\mathbf{p}}\cdot\left(\mathbf{x}-\mathbf{x}'\right)}}{V}\frac{e^{-i\omega\left(t_{x}-t_{x'}\right)}\mathrm{d}\omega}{2\pi},
\end{eqnarray}
with $\varepsilon_{\mathbf{p}}=\left|\mathbf{p}\right|^{2}/\left(2\,m_{N}\right)$. 
\item The ion-electron Coulomb potential propagator between $x$ and $x'$
is
\begin{eqnarray}
iD_{\mathrm{Cou}}\left(x,\,x'\right) & = & \sum_{\mathbf{k}}\int iV_{Ne}\left(\mathbf{k}\right)\frac{e^{i\mathbf{k}\cdot\left(\mathbf{x}-\mathbf{x}'\right)}}{V}\frac{e^{-i\omega\left(t_{x}-t_{x'}\right)}\mathrm{d}\omega}{2\pi}\nonumber \\
 & = & \sum_{\mathbf{k}}i\left(\frac{Z_{\mathrm{ion}}4\pi\alpha}{\left|\mathbf{k}\right|^{2}+i0^{+}}\right)\frac{e^{i\mathbf{k}\cdot\left(\mathbf{x}-\mathbf{x}'\right)}}{V}\delta\left(t_{x}-t_{x'}\right).
\end{eqnarray}
\end{itemize}
In addition, the propagator of electrons in the crystal is similar
to that of the nucleus. It is not shown here because it does not appear
in the Migdal scattering process at the tree level.

On the other hand, we summarize the Feynman rules for the external
legs as the following. 
\begin{itemize}
\item The incoming and outgoing states of the DM particle at space-time
coordinate $x$ are represented with $e^{i\mathbf{p}_{\chi}\cdot\mathbf{x}-i\varepsilon_{\mathbf{p}_{\chi}}t_{x}}/\sqrt{V}$
and $e^{-i\mathbf{p}_{\chi}'\cdot\mathbf{x}+i\varepsilon_{\mathbf{p}_{\chi}'}t_{x}'}/\sqrt{V}$,
respectively, with their corresponding energies $\varepsilon_{\mathbf{p}_{\chi}}=\left|\mathbf{p}_{\chi}\right|^{2}/\left(2\,m_{\chi}\right)$
and $\varepsilon_{\mathbf{p}_{\chi}'}=\left|\mathbf{p}_{\chi}'\right|^{2}/\left(2\,m_{\chi}\right)$. 
\item The incoming and outgoing states of nucleus at $x$ are represented
with $e^{i\mathbf{p}_{N}\cdot\mathbf{x}-i\varepsilon_{\mathbf{p}_{N}}t_{x}}/\sqrt{V}$
and $e^{-i\mathbf{p}_{N}'\cdot\mathbf{x}+i\varepsilon_{\mathbf{p}_{N}'}t_{x}'}/\sqrt{V}$,
respectively, with their corresponding energies $\varepsilon_{\mathbf{p}_{N}}=\left|\mathbf{p}_{N}\right|^{2}/\left(2\,m_{N}\right)$
and $\varepsilon_{\mathbf{p}_{N}'}=\left|\mathbf{p}_{N}'\right|^{2}/\left(2\,m_{N}\right)$. 
\item The incoming and outgoing states of electron in crystal at $x$ are
represented with $\psi_{j}\left(\mathbf{x}\right)e^{-i\varepsilon_{j}t_{x}}$
and $\psi_{i}^{*}\left(\mathbf{x}\right)e^{i\varepsilon_{i}t_{x}}$,
respectively, with their corresponding energies $\varepsilon_{j}$
and $\varepsilon_{i}$. 
\end{itemize}
It is straightforward to translate above Feynman rules in the position
space into those in the momentum space after integrating out the position
4-vector coordinates at each vertex. They are summarized as follows. 
\begin{itemize}
\item Except for the case concerning the external legs of electrons in solids,
each vertex contributes a factor $\left(2\pi\right)V$ and an energy-momentum
conservation condition presented as discrete delta functions $\delta_{\left(\sum_{i}\mathbf{p}_{i}\right),\mathbf{0}}\,\delta\left(\sum\varepsilon_{i}\right)$. 
\item Vertex that contains both the incoming and outgoing states ($\ket{j}$
and $\ket{i}$) of the electrons in solids contributes a factor $\left(2\pi\right)\braket{i|e^{i\mathbf{p}\cdot\hat{\mathbf{x}}}|j}$
and the energy conservation condition $\delta\left(\sum\varepsilon_{i}\right)$,
where $\mathbf{p}$ is the net momentum that sinks into the vertex. 
\item Each DM particle or each nucleus external leg contributes a factor
$1/\sqrt{V}$. 
\item Each internal line corresponds to the propagator of its kind, as well
as a factor $1/V$. For example, the nucleus internal line is read
as $i/\left(\omega-\varepsilon_{\mathbf{\mathbf{p}}}+i0^{+}\right)$
multiplied by a factor $1/V$. 
\end{itemize}
Applying these rules, and summing up all the discrete momenta at each
vertex, the $T$-matrix can be expressed in terms of the amplitude
as the following, 
\begin{eqnarray}
i\mathcal{T} & = & i\mathcal{M}\,\frac{\left(2\pi\right)}{V}\delta_{\left(\sum_{i}\mathbf{p}_{i}\right),\mathbf{0}}\,\delta\left(\sum\varepsilon_{i}\right).
\end{eqnarray}
Above usage of Feynman diagrams and rules in scattering theory can
be transplanted to discussions in the linear response theory in a
parallel fashion, where the imaginary time $t\rightarrow-i\tau$ is
introduced to account for the statistical field theory at a finite
temperature. For example, a fermion propagator in the position space
is written as 
\begin{eqnarray}
S\left(x,\,x'\right) & = & \sum_{i}\sum_{n}\int\frac{u_{i}\left(\mathbf{x}\right)u_{i}^{*}\left(\mathbf{x}'\right)}{i\omega_{n}-\varepsilon_{i}}\frac{e^{-i\omega_{n}\left(\tau-\tau'\right)}}{\beta},
\end{eqnarray}
where $\tau$ and $\tau'$ are the ``temporal'' coordinates, $\beta=1/kT$
is the inverse temperature, with Boltzmann constant $k$, $\omega_{n}=\left(n+\frac{1}{2}\right)\left(2\pi/\beta\right)$
is the Matsubara fermion frequency, with $n$ being an integer to
ensure the anti-periodicity, and $u_{i}\left(\mathbf{x}\right)$ is
the normalized $i$th eigen wavefunction, with its eigen energy $\varepsilon_{i}$,
which can be free or bound state. Base on these propagators, Feynman
rules for finite temperature filed theory can also be summarized and
applied to calculation of physical quantities such as dielectric functions,
and polarizabilities in the context of linear response theory.

\subsection{Dielectric function and random phase approximation (RPA)}

\subsubsection{dielectric function and polarizability}

In a system where there is an external electromagnetic perturbation
$\phi_{\mathrm{ext}}$, the charge is redistributed and gets polarized.
In order to describe the resulting total potential $\phi_{\mathrm{tot}}$
we invoke the inverse dielectric function $\epsilon^{-1}$ in the
following way~\citep{book:17984},
\begin{eqnarray}
\phi_{\mathrm{tot}}\left(\mathbf{x},t\right) & = & \int\mathrm{d}^{4}x'\,\epsilon^{-1}\left(\mathbf{x},t;\mathbf{x}',t'\right)\,\phi_{\mathrm{ext}}\left(\mathbf{x}',t'\right),\nonumber \\
 & = & \int\mathrm{d}^{4}x'\,\left[\delta^{4}\left(x-x'\right)+\int\mathrm{d}^{4}x''\,V^{\mathrm{Cou}}\left(x,\,x''\right)\chi_{\rho\rho}^{\mathrm{r}}\left(\mathbf{x}'',t'';\mathbf{x}',t'\right)\right]\,\phi_{\mathrm{ext}}\left(\mathbf{x}',t'\right)
\end{eqnarray}
where $V^{\mathrm{Cou}}\left(x,\,x''\right)$ is the instantaneous
Coulomb interaction, and the polarizability $\chi_{\rho\rho}^{\mathrm{r}}$
is the density-density correlation function 
\begin{eqnarray}
\chi_{\rho\rho}^{\mathrm{r}}\left(\mathbf{x},t;\mathbf{x}',t'\right) & = & \left(-i\right)\varTheta\left(t-t'\right)\,\left\langle \,\left[\hat{\rho}_{I}\left(\mathbf{x},t\right),\,\hat{\rho}_{I}\left(\mathbf{x}',t'\right)\right]\,\right\rangle \label{eq:charge correlation}
\end{eqnarray}
in the context of linear response theory, with $\left\langle \cdots\right\rangle $
denoting the thermal equilibrium average. In a system with translation-invariance,
the polarizability only depends on the differences of the space-time
coordinates, i.e., $\chi^{\mathrm{r}}\left(\mathbf{x},t;\mathbf{x}',t'\right)=\chi^{\mathrm{r}}\left(\mathbf{x}-\mathbf{x}',t-t';0,0\right)$,
so it is convenient to discuss relevant problems in the momentum-energy
space, where the inverse dielectric function can be expressed as a
product
\begin{eqnarray}
\epsilon^{-1}\left(\mathbf{q},\omega\right) & = & 1+V^{\mathrm{Cou}}\left(\mathbf{q}\right)\chi_{\rho\rho}^{\mathrm{r}}\left(\mathbf{q},\omega\right),\label{eq:inverse dielectric function}
\end{eqnarray}
with $V^{\mathrm{Cou}}\left(\mathbf{q}\right)$ being the reciprocal
counterpart of the Coulomb potential in momentum space. Therefore,
the calculation of the polarizability plays the central role in our
investigation of the response of the solids induced by electromagnetic
perturbation, and the DM particle. While the momentum component of
$\chi_{\rho\rho}^{\mathrm{r}}\left(\mathbf{q},\omega\right)$ can
be obtained by a direct Fourier transformation of Eq.~(\ref{eq:charge correlation}),
deriving the frequency part at a finite temperature needs to be carried
out with imaginary time Green's functions, or the so-called Matsubara
Green's functions. The procedure for the density-density correlation
is briefly summarized as follows. First, one introduces the Matsubara
Green's function, which is defined as
\begin{eqnarray}
\chi_{\rho\rho}\left(\mathbf{q},\tau\right) & = & -\int_{V}\left\langle \,\hat{T}_{\tau}\left[\hat{\rho}_{I}\left(\mathbf{x},\tau\right),\,\hat{\rho}_{I}\left(\mathbf{0},0\right)\right]\,\right\rangle e^{-i\mathbf{q}\cdot\mathbf{x}}\,\mathrm{d}^{3}x\nonumber \\
 & = & \sum_{n}\chi_{\rho\rho}\left(\mathbf{q},i\nu_{n}\right)\frac{e^{-i\nu_{n}\tau}}{\beta},\label{eq:Matsubara Green F}
\end{eqnarray}
where $\hat{T}_{\tau}$ is the imaginary time-ordering operator, $\beta>\tau>0$,
and $\nu_{n}=n\left(2\pi/\beta\right)$ is bosonic frequency since
$\hat{\rho}$ is a bosonic operator, with $n$ being an integer to
ensure the periodicity. Then one obtains $\chi_{\rho\rho}^{\mathrm{r}}\left(i\nu_{n}\right)$
with the Fourier transformation
\begin{eqnarray}
\chi_{\rho\rho}\left(\mathbf{q},i\nu_{n}\right) & = & \int_{0}^{\beta}\chi_{\rho\rho}\left(\mathbf{q},\tau\right)\,e^{i\nu_{n}\tau}\mathrm{d}\tau.\label{eq:reciprocal density-density}
\end{eqnarray}
With the Lehmann representation, it is observed that $\chi_{\rho\rho}^{\mathrm{r}}\left(\mathbf{q},\omega\right)$
and $\chi_{\rho\rho}\left(\mathbf{q},i\nu_{n}\right)$ are just special
case of the same function defined in the entire complex plane except
for a series of poles lying along the real axis. Thus, once $\chi_{\rho\rho}\left(\mathbf{q},i\nu_{n}\right)$
is obtained from Eq.~(\ref{eq:reciprocal density-density}), the
\textit{retarded} polarizability can be derived by performing the
analytic continuation $\chi_{\rho\rho}^{\mathrm{r}}\left(\mathbf{q},\omega\right)=\chi_{\rho\rho}\left(\mathbf{q},i\nu_{n}\rightarrow\omega+i0^{+}\right)$.

\subsubsection{RPA}

\begin{figure*}
\begin{centering}
\includegraphics[scale=0.5]{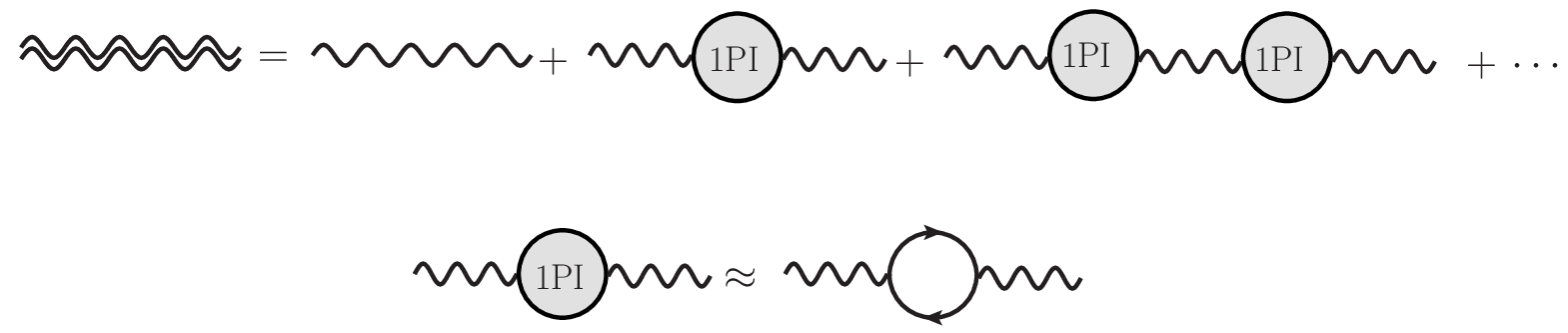}
\par\end{centering}
\caption{\label{fig:Wiggles} \textit{Top}:\textit{ }The diagrammatic expansion
of the Schwinger-Dyson equation for the screened Coulomb interaction,
where the double wiggle line on the left-hand-side of equation represents
the renormalized Coulomb interaction, while the single wiggle represents
a bare Coulomb interaction.\textit{ Bottom}: In the RPA, 1PI diagram
is approximated as an electron-hole loop. See text for details.}
\end{figure*}
The linear response theory can be discussed with our familiar language
of path integrals, as well as necessary modifications accounting for
the imaginary-time argument. The dielectric properties can be described
in terms of diagrams in the top panel of Fig.~\ref{fig:Wiggles},
where the Schwinger-Dyson equation for the screened Coulomb interaction
is represented by a bare interaction wiggly line attached to a geometric
series over polarization bubbles. If we use $W$ and $\Pi$ to represent
the screened Coulomb potential, and the self-energy corresponding
to the 1-particle-irreducible (1PI) blob, respectively, the infinite
sum is read as 
\begin{eqnarray}
W & = & \frac{V^{\mathrm{Cou}}}{\epsilon}\nonumber \\
 & = & V^{\mathrm{Cou}}+V^{\mathrm{Cou}}\left[\Pi+\Pi\,V^{\mathrm{Cou}}\,\Pi+\Pi\,V^{\mathrm{Cou}}\,\Pi\,V^{\mathrm{Cou}}\,\Pi+\cdots\right]V^{\mathrm{Cou}}\nonumber \\
 & = & \frac{V^{\mathrm{Cou}}}{1-V^{\mathrm{Cou}}\,\Pi},
\end{eqnarray}
and hence one has 
\begin{eqnarray}
\epsilon\left(\mathbf{q},i\nu_{n}\right) & = & 1-V^{\mathrm{Cou}}\left(\mathbf{q}\right)\,\Pi\left(\mathbf{q},i\nu_{n}\right).
\end{eqnarray}
On the other hand, adopting the random phase approximation~(RPA)
means that, only the electron-hole bubble is retained among all 1PI
diagrams in calculation of the dielectric functions, which is illustrated
in the bottom panel of Fig.~\ref{fig:Wiggles}. Thus within the framework
of RPA, starting from Eq.~(\ref{eq:Matsubara Green F}) and Eq.~(\ref{eq:reciprocal density-density}),
one first obtains the 1PI blob 
\begin{eqnarray}
\Pi\left(\mathbf{q},i\nu_{n}\right) & \simeq & \frac{1}{V}\sum_{i,j}\frac{\left|\braket{i|e^{i\mathbf{q}\cdot\hat{\mathbf{x}}}|j}\right|^{2}}{\varepsilon_{ij}-i\nu_{n}}\left(n_{i}-n_{j}\right),
\end{eqnarray}
where $n_{i}$~($n_{j}$) denotes the occupation number of the state
$\ket{i}$~($\ket{j}$), and then performing the analytic extension
$i\nu_{n}\rightarrow\omega+i0^{+}$ and inserting the polarizability
into Eq.~(\ref{eq:inverse dielectric function}), one finally arrives
at the Lindhard dielectric function
\begin{eqnarray}
\epsilon^{\mathrm{RPA}}\left(\mathbf{q},\omega\right) & = & 1-\frac{V^{\mathrm{Cou}}\left(\mathbf{q}\right)}{V}\sum_{i,j}\frac{\left|\braket{i|e^{i\mathbf{q}\cdot\hat{\mathbf{x}}}|j}\right|^{2}}{\varepsilon_{ij}-\omega-i0^{+}}\left(n_{i}-n_{j}\right)\label{eq:RPA-dielectric-fucnition0}
\end{eqnarray}
and its inverse 
\begin{eqnarray}
\mathrm{Im}\left[\frac{-1}{\epsilon^{\mathrm{RPA}}\left(\mathbf{q},\omega\right)}\right] & \simeq & \mathrm{Im}\left[-1-\frac{V^{\mathrm{Cou}}\left(\mathbf{q}\right)}{V}\sum_{i,j}\frac{\left|\braket{i|e^{i\mathbf{q}\cdot\hat{\mathbf{x}}}|j}\right|^{2}}{\varepsilon_{ij}-\omega-i0^{+}}\left(n_{i}-n_{j}\right)\right]\nonumber \\
 & = & 2\pi\frac{V^{\mathrm{Cou}}\left(\mathbf{q}\right)}{V}\,\sum_{i,\,j}\left|\braket{i|e^{i\mathbf{q}\cdot\hat{\mathbf{x}}}|j}\right|^{2}\,\delta\left(\varepsilon_{i}-\varepsilon_{j}-\omega\right).
\end{eqnarray}
In above discussion we assume that the electronic system possesses
a translational symmetry, which is true for the case such as homogeneous
electron gas~(HEG). However, for the case of crystal structure where
the translational symmetry for continuous space reduces to that for
the crystal lattice, the polarizability can no longer be expressed
as differences of the space-time coordinates. In this case, the double
periodic position function such as $\chi\left(\mathbf{x},\mathbf{x}';\omega\right)$
can be expressed in the reciprocal space as the following, 
\begin{eqnarray}
\chi\left(\mathbf{x},\mathbf{x}';\omega\right) & = & \frac{1}{V}\sum_{\mathbf{q}\in1\mathrm{BZ}}\sum_{\mathbf{G},\mathbf{G}'}e^{i\left(\mathbf{q}+\mathbf{G}\right)\cdot\mathbf{x}}\,\chi_{\mathbf{G},\mathbf{G}'}\left(\mathbf{q};\omega\right)\,e^{-i\left(\mathbf{q}+\mathbf{G}'\right)\cdot\mathbf{x}'}
\end{eqnarray}
where $\chi_{\mathbf{G},\mathbf{G}'}\left(\mathbf{q};\omega\right)$
is the reciprocal matrix with $\mathbf{G}$ and $\mathbf{G}'$ being
reciprocal lattice vectors and $\mathbf{q}$ is restricted to the
first BZ, which can be determined with the Fourier transformation
\begin{eqnarray}
\chi_{\mathbf{G},\mathbf{G}'}\left(\mathbf{q};\omega\right) & = & \frac{1}{V}\int\mathrm{d}^{3}x\,\mathrm{d}^{3}x\,'e^{-i\left(\mathbf{q}+\mathbf{G}\right)\cdot\mathbf{x}}\chi\left(\mathbf{x},\mathbf{x}';\omega\right)\,e^{i\left(\mathbf{q}+\mathbf{G}'\right)\cdot\mathbf{x}'}.\label{eq:inverse GG}
\end{eqnarray}
As a consequence, for an arbitrary momentum transfer $\mathbf{Q}$,
which can be split into a reduced momentum confined in the 1BZ, and
a reciprocal one, i.e., $\mathbf{\mathbf{Q}}=\mathbf{q}+\mathbf{G}$,
above Lindhard formula relevant for the excitation is generalized
to the following microscopic dielectric matrix:
\begin{eqnarray}
\epsilon_{\mathbf{G},\mathbf{G}'}^{\mathrm{RPA}}\left(\mathbf{q},\omega\right) & = & \delta_{\mathbf{G},\mathbf{G}'}-\frac{V_{\mathbf{G},\mathbf{G}}^{\mathrm{Cou}}\left(\mathbf{q}\right)}{V}\sum_{i,j}\frac{\braket{i|e^{i\left(\mathbf{\mathbf{\mathbf{q}+\mathbf{G}}}'\right)\cdot\hat{\mathbf{x}}}|j}\braket{j|e^{-i\left(\mathbf{\mathbf{\mathbf{q}+\mathbf{G}}}\right)\cdot\hat{\mathbf{x}}}|i}}{\varepsilon_{ij}-\omega-i0^{+}}\left(n_{i}-n_{j}\right)\nonumber \\
 & = & \delta_{\mathbf{G},\mathbf{G}'}-\frac{2\times4\pi\alpha}{V\,\left|\mathbf{q}+\mathbf{G}\right|^{2}}\sum_{i'}^{c}\sum_{i}^{v}\sum_{\mathbf{k},\mathbf{k}'\in1\mathrm{BZ}}\left[-\frac{\braket{i'\,\mathbf{k}'|e^{i\left(\mathbf{\mathbf{\mathbf{q}+\mathbf{G}}}'\right)\cdot\hat{\mathbf{x}}}|i\,\mathbf{k}}\braket{i'\,\mathbf{k}'|e^{i\left(\mathbf{\mathbf{\mathbf{q}+\mathbf{G}}}\right)\cdot\hat{\mathbf{x}}}|i\,\mathbf{k}}^{*}}{\varepsilon_{i'\,\mathbf{k}'}-\varepsilon_{i\,\mathbf{k}}-\omega-i0^{+}}\right]\nonumber \\
 & = & \delta_{\mathbf{G},\mathbf{G}'}+\frac{2\times4\pi\alpha}{V\,\left|\mathbf{q}+\mathbf{G}\right|^{2}}\sum_{i'}^{c}\sum_{i}^{v}\sum_{\mathbf{k}\in1\mathrm{BZ}}\left[\frac{\left(\int_{\Omega}\mathrm{d}^{3}x\,u_{i'\mathbf{k}+\mathbf{q}}^{*}\left(\mathbf{x}\right)\,e^{i\mathbf{\mathbf{G}}'\cdot\mathbf{x}}\,u_{i\mathbf{k}}\left(\mathbf{x}\right)\right)\left(\int_{\Omega}\mathrm{d}^{3}x\,u_{i'\mathbf{k}+\mathbf{q}}^{*}\left(\mathbf{x}\right)\,e^{i\mathbf{\mathbf{G}}\cdot\mathbf{x}}\,u_{i\mathbf{k}}\left(\mathbf{x}\right)\right)^{*}}{\varepsilon_{i'\,\mathbf{k}+\mathbf{q}}-\varepsilon_{i\,\mathbf{k}}-\omega-i0^{+}}\right],\nonumber \\
\label{eq:RPA matrix}
\end{eqnarray}
where $V_{\mathbf{G},\mathbf{G}'}^{\mathrm{Cou}}\left(\mathbf{q}\right)=V^{\mathrm{Cou}}\left(\mathbf{q}+\mathbf{G}\right)\delta_{\mathbf{G},\mathbf{G}'}=4\pi\alpha\,\delta_{\mathbf{G},\mathbf{G}'}/\left|\mathbf{q}+\mathbf{G}\right|^{2}$
is obtained from Eq.~(\ref{eq:inverse GG}), and the inverse dielectric
function can be determined via matrix inversion. 

\subsubsection{\label{sec:cross_section}Cross section and event rate}

Here we first derive a general formula dedicated to the calculation
of two-body scattering cross section for the case of discrete momenta.
By modifying the derivation in textbook, we obtain the cross section
between particles $A$ and $B$,
\begin{eqnarray}
\sigma & = & \left(\prod_{f}\sum_{\mathbf{p}_{f}}\right)\frac{\left|\mathcal{M}\left(\mathbf{k}_{A},\mathbf{k}_{B}\rightarrow\left\{ \mathbf{p}_{f}\right\} \right)\right|^{2}}{\left|\mathbf{v}_{A}-\mathbf{v}_{B}\right|}\frac{\left(2\pi\right)}{V}\,\delta_{\mathbf{k}_{A}+\mathbf{k}_{B},\sum_{_{f}}\mathbf{p}_{f}}\,\delta\left(\varepsilon_{A}+\varepsilon_{B}-\sum_{f}\varepsilon_{f}\right),
\end{eqnarray}
where $\mathbf{v}_{A}-\mathbf{v}_{B}$ is the relative velocity between
the two particles, $\mathbf{p}_{f}$ is the momentum of the $f$th
outgoing particle, and $\mathcal{M}\left(\mathbf{k}_{A},\mathbf{k}_{B}\rightarrow\left\{ \mathbf{p}_{f}\right\} \right)$
is the amplitude in the discrete momentum space. Following the aforementioned
Feynman rules, one can read and approximate the $T$-matrix from Fig.~\ref{fig:Diagrams}
as follows
\begin{eqnarray}
\braket{\mathbf{p}'_{\chi}\mathbf{p}_{N}\mathrm{;electron\,hole\,pair}|iT|\mathbf{p}_{\chi}\mathbf{0}_{N}} & = & i\frac{\left(2\pi\right)}{V}\sum_{\mathbf{k}}\delta_{\mathbf{p}'_{\chi}+\mathbf{p}_{N}+\mathbf{k},\,\mathbf{p}{}_{\chi}}\delta\left(\sum_{i}\varepsilon_{i}\right)\times\left(\cdots\right)_{\mathbf{k}}\nonumber \\
 & \simeq & i\frac{\left(2\pi\right)}{V}\delta_{\mathbf{p}'_{\chi}+\mathbf{p}_{N},\,\mathbf{p}{}_{\chi}}\delta\left(\sum_{i}\varepsilon_{i}\right)\times\sum_{\mathbf{k}}\left(\cdots\right)_{\mathbf{k}}\nonumber \\
 & = & \frac{\left(2\pi\right)}{V}\delta_{\mathbf{p}'_{\chi}+\mathbf{p}_{N},\,\mathbf{p}{}_{\chi}}\delta\left(\sum_{i}\varepsilon_{i}\right)\times i\mathcal{M},
\end{eqnarray}
where in the first line we separate out four-momentum conservation
condition from other terms dependent on transferred momentum $\mathbf{k}$
of the Coulomb potential, and in the second line we assume that $\mathbf{k}$
is much softer than that of the recoiling nucleus $\mathbf{p}_{N}$,
which is consistent with the soft limit condition and explains the
summation over $\mathbf{k}$ in Eq.~(\ref{eq:feynman rules}). Thus,
one obtains the total cross section of an incident DM particle exciting
an electron across the Fermi surface for the HEG via a recoiling nucleus,
\begin{eqnarray}
\sigma & = & \sum_{i,\,j}\sum_{\mathbf{k}',\mathbf{k}}\sum_{\mathbf{\mathbf{p}'_{\chi}}}\sum_{\mathbf{p}_{N}}\frac{2\pi}{V}\int\mathrm{d}\omega\frac{\mathcal{\left|M\right|}^{2}}{v\,\epsilon^{*}\left(\mathbf{k}',\omega\right)\epsilon\left(\mathbf{k},\omega\right)}\delta_{\mathbf{p}'_{\chi}+\mathbf{p}_{N},\,\mathbf{p}{}_{\chi}}\,\delta\left[\frac{p_{\chi}'^{2}}{2m_{\chi}}-\frac{p_{\chi}^{2}}{2m_{\chi}}+\frac{p_{N}^{2}}{2m_{N}}+\varepsilon_{ij}\right]\delta\left(\varepsilon_{ij}-\omega\right)\nonumber \\
 & = & \sum_{\mathbf{k}}\sum_{\mathbf{p}_{N}}\left(\frac{2\pi}{V}\right)\left(\frac{A^{2}\pi\sigma_{\chi n}}{\pi\mu_{\chi n}^{2}\,v}\right)\left(\frac{4\pi\,Z_{\mathrm{ion}}^{2}\,\alpha}{V\,}\right)\int\frac{\mathrm{d}\omega}{\omega^{4}}\,\frac{\left|\mathbf{p}_{N}\cdot\mathbf{\hat{k}}\right|^{2}}{m_{N}^{2}}\,\delta\left[\frac{p_{N}^{2}}{2\mu_{\chi N}}-\mathbf{v}\cdot\mathbf{p}_{N}+\omega\right]\nonumber \\
 &  & \times\frac{1}{\left|\epsilon\left(\mathbf{k},\omega\right)\right|^{2}}\sum_{i,\,j}\frac{2\times4\pi^{2}\alpha}{V\,k^{2}}\left|\braket{i|e^{i\mathbf{k}\cdot\hat{\mathbf{x}}}|j}\right|^{2}\delta\left(\varepsilon_{ij}-\omega\right)\nonumber \\
 & = & \frac{8\pi^{2}A^{2}\sigma_{\chi n}\,Z_{\mathrm{ion}}^{2}\,\alpha}{\mu_{\chi n}^{2}\,v\,}\int\frac{\mathrm{d}^{3}p_{N}}{\left(2\pi\right)^{3}}\int\frac{\mathrm{d}\omega}{\omega^{4}}\,\,\delta\left[\frac{p_{N}^{2}}{2\mu_{\chi N}}-\mathbf{v}\cdot\mathbf{p}_{N}+\omega\right]\int\frac{\mathrm{d}^{3}k}{\left(2\pi\right)^{3}}\frac{\left|\mathbf{p}_{N}\cdot\mathbf{\hat{k}}\right|^{2}}{m_{N}^{2}}\,\mathrm{Im}\left[\frac{-1}{\epsilon\left(\mathbf{k},\omega\right)}\right]\nonumber \\
 & = & \frac{8\pi^{2}A^{2}\sigma_{\chi n}\,Z_{\mathrm{ion}}^{2}\,\alpha}{3\,\mu_{\chi n}^{2}\,v\,\left(2\pi\right)^{6}}\int\left|\mathbf{v}_{\mathrm{ion}}\right|^{2}\,\mathrm{d}^{3}p_{N}\int\frac{\mathrm{d}\omega}{\omega^{4}}\,\,\delta\left[\frac{p_{N}^{2}}{2\mu_{\chi N}}-\mathbf{v}\cdot\mathbf{p}_{N}+\omega\right]\int\mathrm{d}^{3}k\,\mathrm{Im}\left[\frac{-1}{\epsilon\left(\mathbf{k},\omega\right)}\right],\label{eq:sigma}
\end{eqnarray}
where $v$ is the velocity of the DM particle in the frame of laboratory,
$\mathbf{v}_{\mathrm{ion}}=\mathbf{p}_{N}/m_{N}$ is the velocity
of the recoiled nucleus, and $-\mathrm{Im}\left[\epsilon^{-1}\left(\mathbf{k},\omega\right)\right]$
is called the loss function. Since the amplitude $\braket{i|e^{i\mathbf{k}\cdot\hat{\mathbf{x}}}|j}$
uniquely pins down the momentum $\mathbf{k}$~(through the momentum
conservation), only the diagonal terms of momenta $\mathbf{k}$'s
are relevant in above calculation for the HEG. Following Refs.~\citep{Kurinsky_2020,Kozaczuk:2020uzb},
we adopt a Coulomb potential between the recoiled ion and electrons
screened by dielectric function $\epsilon\left(\mathbf{k},\omega\right)$
so as to obtain a general formula that includes self-interaction effects
of the Coulomb propagator. To this end, we take the correspondence
\begin{eqnarray}
2\times\sum_{i,\,j}\frac{4\pi^{2}\alpha}{V\,k^{2}}\left|\braket{i|e^{i\mathbf{k}\cdot\hat{\mathbf{x}}}|j}\right|^{2}\delta\left(\varepsilon_{ij}-\omega\right) & \rightarrow & \mathrm{Im}\left[\epsilon\left(\mathbf{k},\omega\right)\right],
\end{eqnarray}
and use the relation
\begin{eqnarray}
\frac{\mathrm{Im}\left[\epsilon\left(\mathbf{k},\omega\right)\right]}{\left|\epsilon\left(\mathbf{k},\omega\right)\right|^{2}} & = & \mathrm{Im}\left[\frac{-1}{\epsilon\left(\mathbf{k},\omega\right)}\right].
\end{eqnarray}
Thus, from Eq.~(\ref{eq:sigma}) one gives the Migdal exciting event
rate
\begin{eqnarray}
R & = & \frac{\rho_{\chi}}{m_{\chi}}N_{T}\left\langle \sigma v\right\rangle \nonumber \\
 & = & \frac{\rho_{\chi}}{m_{\chi}}\frac{8\pi^{2}A^{2}\sigma_{\chi n}\,Z_{\mathrm{ion}}^{2}\,\alpha\,N_{T}}{3\,\mu_{\chi n}^{2}\,\left(2\pi\right)^{6}}\int\mathrm{d^{3}}v\,f_{\chi}\left(\mathbf{v}\right)\int\left|\mathbf{v}_{\mathrm{ion}}\right|^{2}\,\mathrm{d}^{3}p_{N}\int\frac{\mathrm{d}\omega}{\omega^{4}}\,\,\delta\left[\frac{p_{N}^{2}}{2\mu_{\chi N}}-\mathbf{v}\cdot\mathbf{p}_{N}+\omega\right]\int\mathrm{d}^{3}k\,\mathrm{Im}\left[\frac{-1}{\epsilon\left(\mathbf{k},\omega\right)}\right]\nonumber \\
 & = & \frac{\rho_{\chi}}{m_{\chi}}\frac{2A^{2}\sigma_{\chi n}\,Z_{\mathrm{ion}}^{2}\,\alpha\,N_{T}}{3\,\mu_{\chi n}^{2}\,\varOmega\,m_{N}^{2}}\int\mathrm{d^{3}}v\,\frac{f_{\chi}\left(\mathbf{v}\right)}{v}\,\int p{}_{N}^{3}\,\mathrm{d}p_{N}\int\frac{\mathrm{d}\omega}{\omega^{4}}\,\varTheta\left[v-v_{\mathrm{min}}\left(p_{N},\,\omega\right)\right]\int\frac{\varOmega\,\mathrm{d}^{3}k}{\left(2\pi\right)^{3}}\,\mathrm{Im}\left[\frac{-1}{\epsilon\left(\mathbf{k},\omega\right)}\right],
\end{eqnarray}
where $v_{\mathrm{min}}$ is defined in Eq.~(\ref{eq:Vmin}). In
contrast to the case in HEG, in crystalline environment both the diagonal
and the off-diagonal terms contribute to the total event rate \footnote{{\selectfont See also Ref.~\citep{Knapen:2020aky}.}},
because in this case $\braket{i|e^{i\mathbf{k}\cdot\hat{\mathbf{x}}}|j}$
only constrains momentum $\mathbf{k}$ up to a reciprocal lattice
vector $\mathbf{G}$, and thus the off-diagonal terms $\left(\mathbf{G},\,\mathbf{G}'\right)$
survive when generalizing the above formula:
\begin{eqnarray}
R & = & \frac{\rho_{\chi}}{m_{\chi}}\frac{\pi A^{2}\sigma_{\chi n}\,Z_{\mathrm{ion}}^{2}\,N_{T}}{\mu_{\chi n}^{2}\,\left(2\pi\right)^{3}}\int\mathrm{d^{3}}v\,\frac{f_{\chi}\left(\mathbf{v}\right)}{v}\int\,\frac{\mathrm{d}^{3}p_{N}}{p_{N}}\int\frac{\mathrm{d}\omega}{\omega^{4}}\,\varTheta\left[v-v_{\mathrm{min}}\left(p_{N},\,\omega\right)\right]\nonumber \\
 &  & \times\sum_{\mathbf{G}',\mathbf{G}}\int_{1\mathrm{BZ}}\frac{\mathrm{d}^{3}\left[k\right]}{\left(2\pi\right)^{3}}\,\left[\frac{\mathbf{p}_{N}\cdot\left(\left[\mathbf{k}\right]+\mathbf{G}\right)}{m_{N}}\right]\left[\frac{\mathbf{p}_{N}\cdot\left(\left[\mathbf{k}\right]+\mathbf{G}'\right)}{m_{N}}\right]\mathrm{Im}\left[-\epsilon_{\mathbf{G},\mathbf{G}'}^{-1\mathrm{\left(RPA\right)}}\left(\mathbf{\left[\mathbf{k}\right]},\omega\right)\right]V_{\mathbf{G}',\mathbf{G}'}^{\mathrm{Cou}}\left(\mathbf{\left[\mathbf{k}\right]}\right),
\end{eqnarray}
where $\epsilon^{-1\mathrm{\left(RPA\right)}}$ is the inverse dielectric
matrix introduced in Eq.~(\ref{eq:RPA matrix}). Using the Legendre
addition theorem, one first integrates out the solid angle of nucleus
momentum $\mathbf{p}_{N}$, and then the event rate is recast as
\begin{eqnarray}
R & = & \frac{\rho_{\chi}}{m_{\chi}}\frac{2A^{2}\sigma_{\chi n}\,Z_{\mathrm{ion}}^{2}\,\alpha\,N_{T}}{3\,\mu_{\chi n}^{2}\,\varOmega\,m_{N}^{2}}\int\mathrm{d^{3}}v\,\frac{f_{\chi}\left(\mathbf{v}\right)}{v}\int p{}_{N}^{3}\,\mathrm{d}p_{N}\int\frac{\mathrm{d}\omega}{\omega^{4}}\,\varTheta\left[v-v_{\mathrm{min}}\left(p_{N},\,\omega\right)\right]\nonumber \\
 &  & \times\sum_{\mathbf{G},\mathbf{G}'}\int_{1\mathrm{BZ}}\frac{\varOmega\,\mathrm{d}^{3}\left[k\right]}{\left(2\pi\right)^{3}}\,\frac{\left(\left[\mathbf{k}\right]+\mathbf{G}\right)\cdot\left(\left[\mathbf{k}\right]+\mathbf{G}'\right)}{\left|\left[\mathbf{k}\right]+\mathbf{G}'\right|^{2}}\,\mathrm{Im}\left[-\epsilon_{\mathbf{G},\mathbf{G}'}^{-1\mathrm{\left(RPA\right)}}\left(\mathbf{\left[k\right]},\omega\right)\right].
\end{eqnarray}
In practical implementation of the $\mathtt{YAMBO}$ code, an alternative
dielectric matrix $\widetilde{\epsilon}_{\mathbf{G},\mathbf{G}'}^{\mathrm{\left(RPA\right)}}\left(\mathbf{\left[k\right]},\omega\right)=\left|\left[\mathbf{k}\right]+\mathbf{G}\right|\cdot\epsilon_{\mathbf{G},\mathbf{G}'}^{\mathrm{\left(RPA\right)}}\left(\mathbf{\left[k\right]},\omega\right)\cdot\left|\left[\mathbf{k}\right]+\mathbf{G}'\right|^{-1}$,
or equivalently, $\widetilde{\epsilon}_{\mathbf{G},\mathbf{G}'}^{-1\mathrm{\left(RPA\right)}}\left(\mathbf{\left[k\right]},\omega\right)=\left|\left[\mathbf{k}\right]+\mathbf{G}\right|\cdot\epsilon_{\mathbf{G},\mathbf{G}'}^{-1\mathrm{\left(RPA\right)}}\left(\mathbf{\left[k\right]},\omega\right)\cdot\left|\left[\mathbf{k}\right]+\mathbf{G}'\right|^{-1}$
is preferred, so above event rate is expressed in a symmetrized manner
as follows,
\begin{eqnarray}
R & = & \frac{\rho_{\chi}}{m_{\chi}}\frac{2A^{2}\sigma_{\chi n}\,Z_{\mathrm{ion}}^{2}\,\alpha\,N_{T}}{3\,\mu_{\chi n}^{2}\,\varOmega\,m_{N}^{2}}\int\mathrm{d^{3}}v\,\frac{f_{\chi}\left(\mathbf{v}\right)}{v}\int p{}_{N}^{3}\,\mathrm{d}p_{N}\int\frac{\mathrm{d}\omega}{\omega^{4}}\,\varTheta\left[v-v_{\mathrm{min}}\left(p_{N},\,\omega\right)\right]\nonumber \\
 &  & \times\sum_{\mathbf{G},\mathbf{G}'}\int_{1\mathrm{BZ}}\frac{\varOmega\,\mathrm{d}^{3}\left[k\right]}{\left(2\pi\right)^{3}}\,\frac{\left(\left[\mathbf{k}\right]+\mathbf{G}\right)\cdot\left(\left[\mathbf{k}\right]+\mathbf{G}'\right)}{\left|\left[\mathbf{k}\right]+\mathbf{G}\right|\left|\left[\mathbf{k}\right]+\mathbf{G}'\right|}\,\mathrm{Im}\left[-\widetilde{\epsilon}_{\mathbf{G},\mathbf{G}'}^{-1\mathrm{\left(RPA\right)}}\left(\mathbf{\left[k\right]},\omega\right)\right].
\end{eqnarray}

\twocolumngrid
\renewcommand{\baselinestretch}{1.1}

\bibliographystyle{JHEP1}
\addcontentsline{toc}{section}{\refname}\bibliography{Excitation_Semiconductor}

\end{document}